\documentclass[useAMS,usenatbib]{mn2e}

\usepackage{txfonts}
\usepackage[authoryear]{natbib}
\usepackage{lscape}                                
\usepackage{color}
\usepackage{graphicx}
\usepackage{longtable}

\title[\emph{s}-process enrichment in NGC~3918]{\emph{s}-process enrichment in the planetary nebula NGC~3918. Results from deep echelle spectrophotometry \thanks{Based on observations collected at the European Southern Observatory, Chile, proposal number ESO 090.D-0265(A)}}

\author[J. Garc\'{\i}a-Rojas et al.]
{J.~Garc\'{\i}a-Rojas$^{1,2}$\thanks{E-mail: jogarcia@iac.es}
S. Madonna$^{1,2}$, V. Luridiana $^{1,2}$, N.~C. Sterling$^3$, C. Morisset$^4$, 
\newauthor
G. Delgado-Inglada$^4$, L. Toribio San Cipriano$^{1,2}$ \\
$^1$Instituto de Astrof\'\i sica de Canarias, E-38205 La Laguna, Tenerife, Spain \\
$^{2}$Universidad de La Laguna, Dpto. Astrof\'isica,  E-38206 
           La Laguna, Tenerife, Spain\\
$^{3}$ Department of Physics, University of West Georgia, 1601 Maple Street, Carrollton, GA 30118, USA \\
$^{4}$ Instituto de Astronom\'ia, Universidad Nacional Aut\'onoma de M\'exico,
Apdo. Postal 70264, M\'exico D. F., 04510 Mexico}

\newcommand{\elecd}{$n_{\rm e}$}
\newcommand{\te}{$T_{\rm e}$}
\newcommand{\hb}{H$\beta$}
\newcommand{\ha}{H$\alpha$}
\newcommand{\fci}{[C~{\sc i}]}

\newcommand{\foi}{[O~{\sc i}]}
\newcommand{\foii}{[O~{\sc ii}]}
\newcommand{\foiii}{[O~{\sc iii}]}

\newcommand{\fsii}{[S~{\sc ii}]}
\newcommand{\fsiii}{[S~{\sc iii}]}
\newcommand{\fnitroi}{[N~{\sc i}]}
\newcommand{\fnii}{[N~{\sc ii}]}
\newcommand{\sfmgi}{Mg~{\sc i}]}
\newcommand{\mgi}{Mg~{\sc i}}
\newcommand{\mgii}{Mg~{\sc ii}}
\newcommand{\fariii}{[Ar~{\sc iii}]}
\newcommand{\fariv}{[Ar~{\sc iv}]}
\newcommand{\farv}{[Ar~{\sc v}]}

\newcommand{\fclii}{[Cl~{\sc ii}]}
\newcommand{\fcliii}{[Cl~{\sc iii}]}
\newcommand{\fcliv}{[Cl~{\sc iv}]}
\newcommand{\fcrii}{[Cr~{\sc ii}]}
\newcommand{\fcriii}{[Cr~{\sc iii}]}

\newcommand{\fneiii}{[Ne~{\sc iii}]}
\newcommand{\fneiv}{[Ne~{\sc iv}]}
\newcommand{\fnev}{[Ne~{\sc v}]}
\newcommand{\fkriii}{[Kr~{\sc iii}]}
\newcommand{\fkriv}{[Kr~{\sc iv}]}
\newcommand{\fkrv}{[Kr~{\sc v}]}
\newcommand{\fkiv}{[K~{\sc iv}]}
\newcommand{\fkv}{[K~{\sc v}]}
\newcommand{\fkvi}{[K~{\sc vi}]}
\newcommand{\fxeiii}{[Xe~{\sc iii}]}
\newcommand{\fxeiv}{[Xe~{\sc iv}]}
\newcommand{\fxevi}{[Xe~{\sc vi}]}
\newcommand{\frbiv}{[Rb~{\sc iv}]}
\newcommand{\frbv}{[Rb~{\sc v}]}
\newcommand{\fnaiv}{[Na~{\sc iv}]}

\newcommand{\fniqv}{[Ni~{\sc v}]}
\newcommand{\ffiv}{[F~{\sc iv}]}
\newcommand{\fcuv}{[Cu~{\sc v}]}
\newcommand{\fcuvi}{[Cu~{\sc vi}]}
\newcommand{\fcoiv}{[Co~{\sc iv}]}
\newcommand{\fcovi}{[Co~{\sc vi}]}
\newcommand{\fcav}{[Ca~{\sc v}]}
\newcommand{\fmniii}{[Mn~{\sc iii}]}
\newcommand{\fmniv}{[Mn~{\sc iv}]}
\newcommand{\fmnv}{[Mn~{\sc v}]}
\newcommand{\fmnvi}{[Mn~{\sc vi}]}

\newcommand{\ffeiii}{[Fe~{\sc iii}]}
\newcommand{\ffeiv}{[Fe~{\sc iv}]}
\newcommand{\ffev}{[Fe~{\sc v}]}
\newcommand{\ffevi}{[Fe~{\sc vi}]}
\newcommand{\ffevii}{[Fe~{\sc vii}]}
\newcommand{\fpii}{[P~{\sc ii}]}
\newcommand{\piv}{P~{\sc iv}}

\newcommand{\fseiii}{[Se~{\sc iii}]}
\newcommand{\oiii}{O~{\sc iii}}
\newcommand{\oiv}{O~{\sc iv}}
\newcommand{\ov}{O~{\sc v}}

\newcommand{\nii}{N~{\sc ii}}
\newcommand{\niii}{N~{\sc iii}}
\newcommand{\niv}{N~{\sc iv}}
\newcommand{\nv}{N~{\sc v}}

\newcommand{\silii}{Si~{\sc ii}}
\newcommand{\siliii}{Si~{\sc iii}}
\newcommand{\siliv}{Si~{\sc iv}}
\newcommand{\oi}{O~{\sc i}}
\newcommand{\oii}{O~{\sc ii}}
\newcommand{\ci}{C~{\sc i}}
\newcommand{\cii}{C~{\sc ii}}
\newcommand{\ciii}{C~{\sc iii}}
\newcommand{\civ}{C~{\sc iv}}

\newcommand{\neii}{Ne~{\sc ii}}
\newcommand{\neiii}{Ne~{\sc iii}}

\newcommand{\nev}{Ne~{\sc v}}
\newcommand{\sii}{S~{\sc ii}}

\newcommand{\hi}{H\,{\sc i}}
\newcommand{\hii}{H~{\sc ii}}

\newcommand{\hei}{He~{\sc i}}
\newcommand{\heii}{He~{\sc ii}}
\newcommand{\mc}{\multicolumn}

\begin{document}

\date{Accepted 2015 June 23. Received 2015 June 22; in original form 2015 March 19}

\pagerange{\pageref{firstpage}--\pageref{lastpage}} \pubyear{2015}

\maketitle

\label{firstpage}

\begin{abstract}
The chemical content of the planetary nebula NGC\,3918 is investigated through deep, high-resolution (R$\sim$40000) UVES at VLT spectrophotometric data. We identify and measure more than 750 emission lines, making ours one of the deepest spectra ever taken for a planetary nebula. Among these lines we detect very faint lines of several neutron-capture elements (Se, Kr, Rb,
and Xe), which enable us to compute their chemical abundances with unprecedented accuracy,
thus constraining the efficiency of the {\emph s}-process and convective dredge-up in NGC\,3918’s progenitor star. We find that Kr is strongly enriched in NGC\,3918 and that Se is less enriched than Kr, in agreement with the results of previous papers and with predicted {\emph s}-process nucleosynthesis. We also find that Xe is not as enriched by the {\emph s}-process in NGC\,3918 as is Kr and, therefore, that neutron exposure is typical of modestly subsolar metallicity AGB stars. A clear correlation is found when representing [Kr/O] $vs.$ log(C/O) for NGC\,3918 and other objects with detection of multiple ions of Kr in optical data, confirming that carbon is brought to the surface of AGB stars along with {\emph s}-processed material during third dredge-up episodes, as predicted by nucleosynthesis models. We also detect numerous refractory element lines (Ca, K, Cr, Mn, Fe, Co, Ni, and Cu) and a large number of metal recombination lines of C, N, O, and Ne. We compute physical conditions from a large number of diagnostics, which are highly consistent among themselves assuming a three-zone ionization scheme. Thanks to the high ionization of NGC\,3918 we detect a large number of recombination lines of multiple ionization stages of C, N, O and Ne. The abundances obtained for these elements by using recently-determined state-of-the-art ICF schemes or simply adding ionic abundances are in very good agreement, demonstrating the quality of the recent ICF scheme for high ionization planetary nebulae. 
\end{abstract}


\section{Introduction}\label{intro}

About half of the heavy elements ($Z>30$) in the Universe are formed by slow neutron(n)-capture  nucleosynthesis (the ``\emph{s}-process'') in the asymptotic giant branch (AGB) phase, when neutrons are released in the intershell region between the H- and He-burning shells through the  $^{13}$C($\alpha,n$)$^{16}$O reaction or, in more massive AGB stars (M~$>4$~M$_{\odot}$), the $^{22}$Ne($\alpha,n$)$^{25}$Mg reaction. Fe-peak nuclei in this layer, exposed to the neutron flux, experience \emph{n}-captures interlaced with $\beta$-decays that transform them into heavier elements, in the process known as the "\emph{s}-process". The \emph{s}-process- and He-burning-enriched material is later conveyed to the stellar envelope by convective dredge-up and released into the interstellar medium by stellar winds and planetary nebula (PN) ejection \citep{bussoetal99, herwig05, kappeleretal11, karakaslattanzio14}, where it can 
eventually be incorporated into a new generation of stars. 

The exact details of this scenario are still poorly understood 
due to the lack of observational constraints to models of AGB nucleosynthesis \citep{karakasetal09}.
Two major problems are strong obscuration by dust, which hinders observations in the optical, and the still-evolving surface composition, which complicates the interpretation of data.
Fortunately, \emph{n}-capture elements can also be studied through nebular spectroscopy of PNe.  PNe are excellent laboratories for such investigations: they are easily observed in the optical region, which is home to a multitude of \emph{s}-process element transitions, including lines from multiple ions of Br, Kr, Rb, and Xe; in addition, nucleosynthesis and dredge-up is complete in these objects, whereas the composition of AGB stars is evolving. 
On the downside, the emission lines of \emph{s}-elements are intrinsically very weak even in the brightest PNe, 
so that deep, high-resolution spectroscopy is required to detect them. Because of these stringent technical requirements, only a few detailed abundance analysis have been published so far
\citep[see, e.g.][]{pequignotbaluteau94, sharpeeetal07, sterlingdinerstein08}.
The determination of total abundances is often limited by the fact that only one ion of each element is detected, leading to large and uncertain corrections for unobserved ions. 

To improve the accuracy of \emph{s}-element abundance determinations in PNe, we embarked on an ambitious observational program aimed at detecting multiple ions of several {\emph n}-capture species in a small sample of about 8 PNe. The data gathered will enable us to compute accurate total abundances for the object of the sample, as well as verify the consistency of current ICFs based on photoionization modelling for future, less in-depth studies. In addition, the comparison of different lines of the same ionization species will enable us to assess the quality of the (as yet poorly-tested) atomic data for heavy elements. Both aspects are crucial to hone nebular spectroscopy into an effective tool for studying \emph{s}-process nucleosynthesis.

Two further scientific goals depend crucially on the availability of a sample of several PNe. On one side, we intend to study the correlation, predicted by current AGB models, between the C/O ratio and the \emph{s}-process enrichment factors. Additionally, we want to explore the correlation between the pattern of \emph{s}-element abundances and the mass of the progenitor star, which, according to theory, is modulated by the nuclear reaction activated in each mass range \citep{vanraaietal12, karakasetal12}. 

Given the deep, high-resolution spectroscopy required to detect optical {\emph n}-capture element
emission lines, such studies naturally reveal numerous weak features of other species. We
have detected more than 750 emission lines in NGC\,3918, making ours one of the deepest
spectra of a PN ever obtained at such a high spectral resolution. These include forbidden lines of several iron-group and other
refractory elements, a multitude of permitted features, and a host of diagnostic lines. We
determine abundances for all species for which atomic data are available in addition to {\emph n}-capture
elements. For comparison, available very deep PNe spectra at a comparable high-resolution in the literature are those of e.~g. \citet{sharpeeetal03} who detected $\sim$800 lines in the PN IC\,418, and \citet{sharpeeetal07} who detected between $\sim$600 and $\sim$900 lines in 4 Galactic PNe. 

In this paper, we describe the results for the first PN of our sample, NGC~3918. The paper is organized as follows: in Sect.~\ref{obsred} we describe the observations and the data reduction; in Sect.~\ref{lineint} we present the table of identified lines as well as the reddening correction; in Sect.~\ref{physcond} and ~\ref{chemabun} we compute the physical conditions and the ionic and total chemical abundances. Finally, in Sect.~\ref{discuss} we discuss our results and draw some conclusions. 
The work is summarized in Sect.~\ref{conclu}.

\section{Observations and Data Reduction}\label{obsred}

The spectra of NGC\,3918 were taken with the Ultraviolet-Visual Echelle Spectrograph \citep[UVES, ][]{dodoricoetal00}, attached to the 8.2m Kueyen (UT2) Very Large Telescope at Cerro Paranal Observatory (Chile) in service mode on 2013 March 8. The observations were taken under clear/dark conditions and and the seeing remains below 1.5$''$ during the whole run (see Table~\ref{tobs}).

\begin{figure}
\begin{center}
\includegraphics[width=9cm]{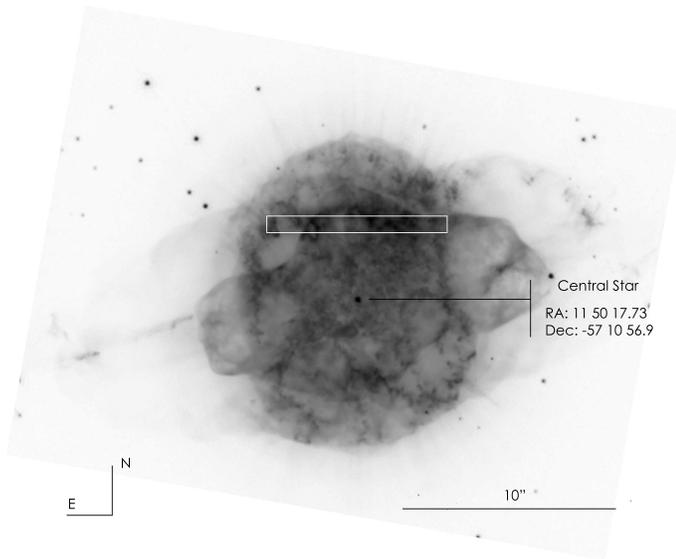}
\caption{Deep HST H$\alpha$ image of NGC\,3918. The position of the central star and its coordinates are shown. 
The slit position is indicated as a white rectangular box. The slit size is 10$''\times1''$. }
\label{slitpos}
\end{center}
\end{figure}

We used two standard settings, DIC1 (346+580) and DIC2 (437+860), in both the red and blue arms of the telescope, covering nearly the full optical range between 3100--10420 \AA. In the setting DIC1 (346+580) the dichroic splits the light beam in two wavelengths ranges: 3100--3885 \AA\ in the blue arm and 4785--6805 \AA\ in the red arm; in the setting DIC2 (437+860) the dichroic configuration change to split the light beam in the wavelength ranges: 3750--4995 in the blue arm and 6700--10420 \AA\ in the red arm. The wavelength regions 5773--5833  \AA\ and 8540--8650  \AA\ were not observed because of the gap between the two CCDs used in the red arm. Additionally, there are small gaps at the reddest wavelengths which were not observed because the redmost orders do not fit completely within the CCD. The journal of observations is shown in Table~\ref{tobs}. The atmospheric dispersion corrector (ADC) was used to compensate for atmospheric dispersion at the large airmasses the object was observed (between 1.5 and 1.7).
The spectra are divided in four spectral ranges (B1, B2, R1 and R2; see Table~\ref{tobs}) because the 
central wavelengths of the two arms were set to two different values to cover the whole optical-NIR range. We obtained 6 exposures of 850 s each in each configuration that were taken consecutively following the sequence DIC1 (346+580) $\rightarrow$ DIC2 (437+860), giving a total exposure time of 1.42 h in each configuration. We have to emphasize that at the end of the observing period, only 42.8\% of the total observing time requested was completed. Fortunately, the spectrum was deep enough to reach part of our scientific goals (detection of faint \emph{n}-capture element lines). Additional single short exposures of 60 s each were taken to obtain non-saturated flux measurements for the brightest emission lines.
The slit length was fixed to 10$''$ in the two bluest spectral ranges (B1 and B2) and 12$''$ in the two reddest ones (R1 and R2), obtaining an adequate interorder separation. The slit width was set to 1$''$, which gives an effective spectral resolution of $\Delta\lambda/\lambda$$\sim$ 40,000 (6.5 km s$^{-1}$), which is needed to deblend some important neutron-capture emission lines from other spectral features \citep{sharpeeetal07}. The final one-dimensional spectra analysed in this paper cover an area of 9.35 arcsec$^2$ common to all spectral ranges.
In Figure~\ref{slitpos} we show a high spatial resolution H$\alpha$ image from the HST archive of NGC\,3918. The slit center was set 3.8$''$ north to the central star of NGC\,3918 oriented E-W (PA=90$^{\circ}$), covering the brightest area of NGC\,3918. 

\setcounter{table}{0}
\begin{table}
\begin{minipage}{75mm}
\centering
\caption{Journal of observations.}
\label{tobs}
\begin{tabular}{c@{\hspace{2.8mm}}c@{\hspace{2.8mm}}c@{\hspace{2.8mm}}c@{\hspace{1.8mm}}c@{\hspace{1.8mm}}}
\noalign{\hrule} \noalign{\vskip3pt}
Telescope& Date & $\Delta\lambda$~(\AA) & Exp. time (s) & Seeing ($''$) \\
\noalign{\vskip3pt} \noalign{\hrule} \noalign{\vskip3pt}
8.2 m VLT& 2013/03/08 & B1: 3100$-$3885 & 60, 6$\times$850 &1.0 --1.4 \\
"&" & B2: 3750$-$4995 & 60, 6$\times$850 & 1.0--1.2 \\
"&" & R1: 4785$-$6805 & 60, 6$\times$850 &  1.0--1.4\\
"&" & R2: 6700$-$10420 & 60, 6$\times$850 &  1.0--1.2\\
\noalign{\vskip3pt} \noalign{\hrule} \noalign{\vskip3pt}
\end{tabular}
\end{minipage}
\end{table}

The raw frames were reduced using the public ESO UVES pipeline \citep{ballesteretal00} under the GASGANO graphic user interface, following the standard procedure of bias subtraction, aperture
extraction, background subtraction, flat-fielding and wavelength calibration. The final products of the pipeline were 2D wavelength calibrated spectra; our own IDL scripts were used thereafter to collapse the spectra in the spatial direction and obtain our final 1D-spectra. 
The standard star LTT 3218 \citep{hamuyetal92, hamuyetal94} was observed to perform the flux
calibration and was also fully reduced with the pipeline. The flux calibration and radial velocity correction were performed using the standard procedures with IRAF\footnote{IRAF is distributed by National Optical Astronomy Observatory, which is operated by Association of Universities for Research in Astronomy, under cooperative agreement with the National Science Foundation.} \citep{tody93}.

\section{Line fluxes, identifications, and extinction correction}\label{lineint}

\begin{figure*} 
\begin{center}
\includegraphics[width=\textwidth]{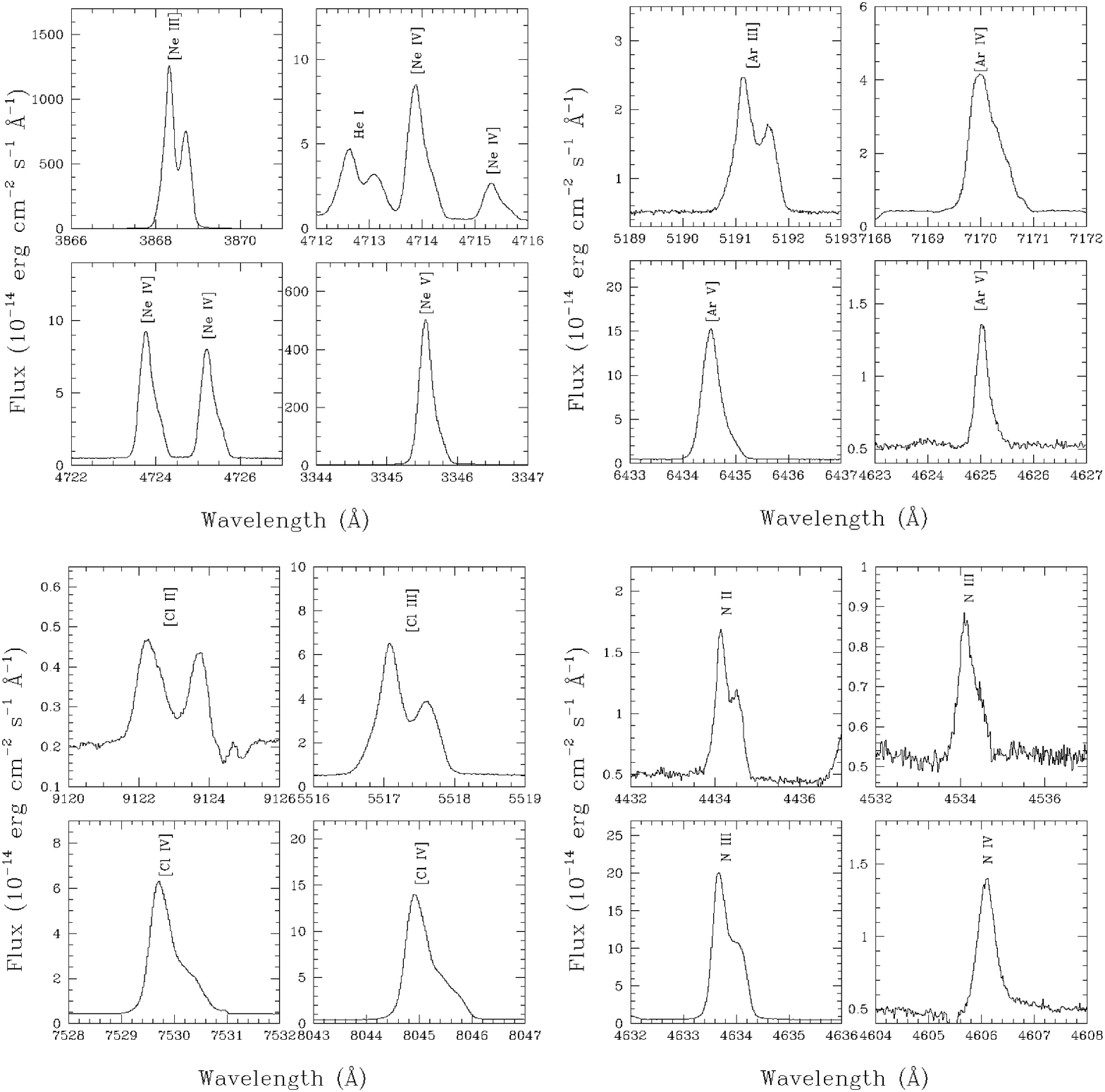}
\caption{Portions of the echelle spectrum of NGC\,3918 showing selected collisionally escited lines of different ionic species of Ne, Ar, and Cl, and recombination lines of N. The effect of the expansion velocity field is clearly shown. As expected, the effect is larger in the lowest ionization species whose lines are formed in the external parts of the PN, than in the highest ionization species, which form in the inner parts of the PN.}
\label{profiles}
\end{center}
\end{figure*}

We used the {\it splot} routine of the IRAF package to measure the line intensities. The expansion velocity field of the PN is mapped in the line profile pattern, which evolves from the double-peaked line profile of the lowest ionization species to the simpler profile of the highest ionization species. Hence, the total flux of each line was measured by integrating the profile between two given limits, over a local continuum estimated by eye. Owing to the small area covered by our slit, we could not extract a sky 
spectrum. However, taking into account the peculiar profile of the emission lines in each range of ionization, it was 
easy to distinguish telluric emission features from nebular emission lines. The cases 
in which nebular emission lines are severely blended with sky emission features are labelled in the table of 
line identifications (Table~\ref{lineid}). Finally, several lines are strongly affected by atmospheric features in absorption or by internal reflections by charge transfer in the CCD, rendering their intensities unreliable. In some cases, where we consider we could deblend the line from the non-nebular feature, we decided to report the line flux anyway, and included a label in the line identification table as a note of caution. 
In Fig.~\ref{profiles}, we present an example of the line profiles for different ionic species of several elements to show the effect of the expansion velocity field on the line profiles.

The four different spectral ranges covered in the spectra have overlapping regions at the edges. To produce a homogeneous data set of line flux ratios, all the lines fluxes of a given spectrum were normalized to a particular non-saturated bright emission line located in such overlapping region. The selected lines were H9 $\lambda$3835 for the B1 and B2 spectra, {\fsii} 6731 \AA\ for R1 and R2, and {\hb} for B2 and R1. All line fluxes were eventually referred to {\hb}.
Some lines that were saturated in the long exposures were measured in the short ones and rescaled to the {\hb} flux in a similar way. 

The final intensity of a given line in the overlapping regions is the average of the values obtained in both spectra. The differences in the fluxes measured for each line in overlapping spectra do not show any systematic trend and, for the brightest lines, are always lower than 10\%. The differences found for the faintest lines in the overlapping regions can be slightly  larger ($\sim$20-25\%), and the final adopted errors for each line take into account these uncertainties. Therefore, the final adopted uncertainties are always larger than the differences found between both ranges.
These differences are probably caused by the fact that these lines are in the red and blue extremes of the CCDs, where the 
flat-field correction might be less reliable. 
We do not expect our final results to be substantially affected by these effects because only line ratios are used in our analysis. 

Thanks to the high ionization of NGC\,3918, we detect multiple lines of different excitation stages. More than 750 emission lines were measured; most of the lines are permitted lines of {\hi}, {\hei} and {\heii}, but there is also a large number of permitted lines of heavier ions, such as {\oii}, {\oiii}, {\oiv}, {\nii}, {\niii}, {\niv}, {\cii}, 
{\ciii}, {\civ}, {\neii}, {\neiii}, {\silii}, {\siliii}, {\mgi} and {\mgii} (and, possibly, {\nv}, {\ov} and {\sii}, see Table~\ref{lineid}).  We also detect several forbidden and semi-forbidden lines from ions such as {\fnitroi}, {\fnii}, {\foi}, {\foii}, {\foiii}, {\ffiv}, {\fneiii}, {\fneiv}, {\fnev}, {\fnaiv}, {\sfmgi}, {\fpii}, {\fsii}, {\fsiii}, {\fclii}, {\fcliii}, {\fcliv}, {\fariii}, {\fariv}, {\farv}, {\fkiv}, {\fkv}, {\fkvi}, {\fcav}, {\fcrii}, {\fcriii}, {\fmniv}, {\fmnv},  {\fmnvi},{\ffeiii}, {\ffeiv}, {\ffev}, {\ffevi}, {\ffevii}, {\fcoiv}, {\fniqv}, {\fcuv}, {\fcuvi}, {\fkriii}, {\fkriv}, {\fkrv}, {\frbiv}, {\frbv}, {\fxeiii} and {\fxeiv} (and, tentatively, {\fcovi}, {\fseiii}, and {\fxevi}). The depth of our spectra allows us to detect lines as fainter as 10$^{-5}$$\times$$I$({\hb}). The identification and adopted laboratory wavelengths of the lines are based on several previous identifications in the literature \citep[see][and references therein]{baluteauetal95, cleggetal87, fangliu11, garciarojasetal09, garciarojasetal12, pequignotbaluteau94, sharpeeetal07, zhangetal05}. We also made use of Peter van Hoof's atomic line list v2.05B18\footnote{http://www.pa.uky.edu/$\sim$peter/newpage/}. Details on the identification of neutron-capture element emission lines are given in Sect.~\ref{s_lines}.

For the reddening correction, we assumed the standard extinction law for the Milky Way parametrized by \citet{seaton79},  
with $R_v$=3.1. We selected this parametrization of the extinction law to be consistent with the analysis of \citet{cleggetal87} which is the deepest study of this object in the optical range. The use of other parametrizations of the standard extinction law would have no significant effects on the results of this paper, as the obtained c({\hb}) is fairly low (see below). The logarithmic redddening coefficient, c(H$\beta$), was derived by fitting the observed 
Balmer lines, $F$({\hi} Balmer)/$F$({\hb}) (from H25 to H3), and the observed $F$({\hi} Paschen)/$F$({\hb})  lines (from P25 to P9) to the theoretical values computed by \citet{storeyhummer95} for {\te}=11000 K and {\elecd}=5000 cm$^{-3}$, as derived for NGC\,3918 
by \citet{cleggetal87}. We only used those lines neither contaminated by telluric emission/absorptions nor by other nebular emissions. We obtained an averaged value of c(H$\beta$)=0.26$\pm$0.06. This value is lower than those obtained by \citet{tsamisetal03b}, \citet{cleggetal87} and \citet{penatorrespeimbert85}, which amount to 0.44, 0.43 and 0.40, respectively. However, our value is consistent with the value obtained by \citet{cleggetal87} from the Balmer decrement, which amounts to 0.33$\pm$0.14, and with the value obtained by \citet{torrespeimbertpeimbert77} (c(H$\beta$)=0.30).

To determine the line flux uncertainties, we considered individually each spectral range (B1, B2, R1, and R2). Several lines were chosen in each of these ranges so as to cover the whole intensity ranges, and the statistical errors associated to the uncertainties in the continuum measurement were computed using the IRAF {\it splot} task.
Error propagation and a logarithmic interpolation of $F$($\lambda$)/$F$(H$\beta$) \emph{ vs.} 
$\sigma$($F$($\lambda$)/$F$(H$\beta$)) 
were used to determine $\sigma$($F$($\lambda$)/$F$(H$\beta$)) for each line. Finally, the final percentile errors (1$\sigma$) of the 
$I(\lambda$)/$I({\rm H}\beta)$ ratios were computed taking into account the 
uncertainties in the determination of
$c$(H$\beta$). The result is presented in column 8 of 
Table~\ref{lineid}. Colons indicate errors higher than 40\%.

\subsection{Unidentified lines.}\label{no_ident}

There are 18 lines that could not be identified with the available sources, amounting to $\sim$2.5\% of the measured lines. There are also about 30 dubious identifications, labelled in Table~\ref{lineid} with a quotation mark. We checked that these lines are not telluric lines, nor ghosts or charge transfer features. Given the strong dependence of the line profiles (velocity field) with the excitation of the ion, we can at least tentatively classify these lines as low or high ionization ions, with low ionization
species having a clear double-peaked line profile and high ionization lines a single-peaked
or slightly double-peaked profile. In Table~\ref{tab_no_ident} we show the list of unidentified lines with a rough estimation of the laboratory wavelength and the profile type.

\setcounter{table}{2}
\begin{table}
\centering
\caption{Unidentified lines and profile type.}
\label{tab_no_ident}
\begin{tabular}{lcc}
\hline
$\lambda_{obs}$  &  Expected $\lambda_0$  & Profile Type$^{\rm a}$  \\
\hline
3571.09	&	$\sim$3571.3	&	low	 \\		
4554.64	&	$\sim$4555.0	&	high	 \\
5150.60	&	$\sim$5151.0	&	high	  \\
5450.62	&	$\sim$5451.0	&	high	  \\
5464.91	&	$\sim$5465.3	&	high	  \\
5467.17	&	$\sim$5467.6	&	low?	  \\
5468.35	&	$\sim$5468.8	&	high	  \\
5487.87	&	$\sim$5488.3	&	high	  \\
5659.80	&	$\sim$5660.3	&	high	  \\
6114.12	&	$\sim$6114.6	&	low-high?   \\
6272.78	&	$\sim$6273.3	&	high	  \\
6958.81	&	$\sim$6959.3	&	high	  \\
7030.44	&	$\sim$7031.0	&	high	  \\
7046.36	&	$\sim$7046.9	&	high	  \\
7058.79	&	$\sim$7059.3	&	high	  \\
7102.30     &       $\sim$7102.8    &       low     \\
7509.38	&	$\sim$7509.9	&	low	  \\
8703.81	&	$\sim$8704.3	&	low 	  \\
\hline
\end{tabular}
\begin{description}
\item[$^{\rm a}$]``Low'' means a double-peak line profile; ``high'' means single-peaked or slightly double-peak line profile (see text).
\end{description}
\end{table}

\subsection{Identification of neutron-capture lines}\label{s_lines}

The detection of neutron-capture element emission lines in the optical spectra of photoionized nebulae is a hard task, since these lines are intrinsically very faint. However, since the first identification of neutron-capture elements lines twenty years ago by \citet{pequignotbaluteau94} in the optical spectrum of the PN NGC\,7027, the reported detections of these lines in the optical spectra of Galactic PNe and {\hii} regions have grown significantly \citep[e.~g.][]{baldwinetal00, yliuetal04a, fangliu11, garciarojasetal09, garciarojasetal12, otsukaetal10, otsukaetal11, otsukatajitsu13, peimbertetal04, sharpeeetal07, sterlingetal09, zhangetal05}. Other studies have reported the detection of neutron-capture element lines in the near-infrared spectra of Galactic PNe and {\hii} regions \citep[e.~g.][]{dinerstein01, sterlingdinerstein08, blummcgregor08} and in other galaxies \citep[e.~g.][]{vanzietal08}. 

\begin{figure}
\begin{center}
\includegraphics[width=\columnwidth]{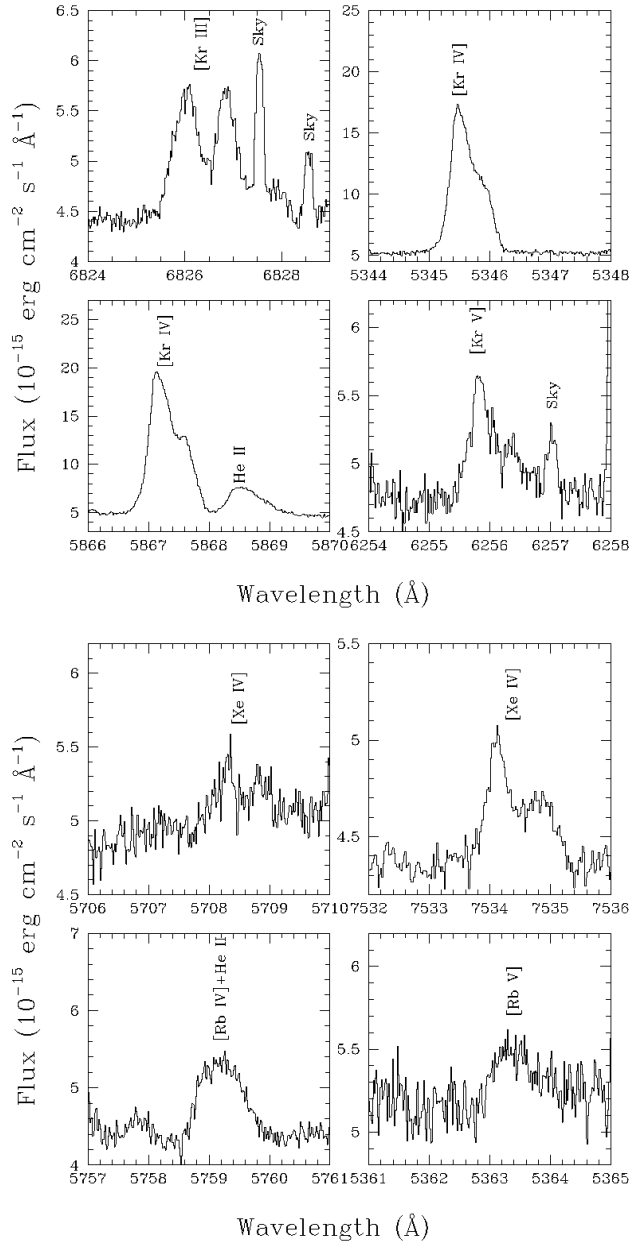}
\caption{Portions of the echelle spectrum of NGC\,3918 showing lines of different ionic 
species of Kr (upper panel), Xe and Rb (lower panel).}
\label{selem_prof}
\end{center}
\end{figure}

We report the detection of several neutron-capture element lines in NGC\,3918. In our analysis, we use all the certain and probable line identifications by \citet{pequignotbaluteau94} and \citet{sharpeeetal07} as a reference. 
The complex velocity structure of NGC\,3918 is a double-edged sword. On the one hand, it does not allow us to easily use automatic identification algorithms such as EMILI \citep{sharpeeetal03}; on the other hand, it helps us to detect the presence of line blends with ions with different ionization potentials. Additionally, the large number of ghosts and charge-transfer features, as well as the ``wiggling'' of the continuum which is typical of deep echelle spectra, do not allow us to properly use the PySSN spectrum synthesis code, except in some cases (see below). The PySSN code is a python version of the X-SSN code \citep{pequignotetal12}  and is still in development. PySSN uses a database of emission lines containing identifications, wavelengths and relative intensities for lines of the same multiplets (or emitted by the same process as atmospheric emission and absorption lines). It generates a synthetic spectrum by summing the contributions of all the individual lines, taking into account different profiles for different ions and/or emission processes, and the continuous nebular emission. Reddening and instrumental response are also taken into account.
Therefore, in most of the cases, we identify the lines by eye, following several criteria: i) the line profiles agree with what is expected for the ionization potential of the ion; in some cases, this criterion allows us to discard possible blends with other faint lines; ii) the radial velocities of different lines of the same ion agree; iii) the relative intensities of lines arising from the same ion agree with what is expected from the collision strengths and spontaneous emission coefficient calculations for these transitions (this applies to the {\fkriv} and {\fxeiv} lines); iv) finally, in some cases of very tight blends we use the PySSN spectrum synthesis code, which makes use of all the previous conditions to perform the line fitting. When the first three criteria are fullfilled we consider the identification to be robust. In Fig.~\ref{selem_prof} we show some of the neutron-capture element lines detected in our spectrum. 

\subsubsection{Se line identifications}

\begin{figure}
\begin{center}
\includegraphics[width=\columnwidth]{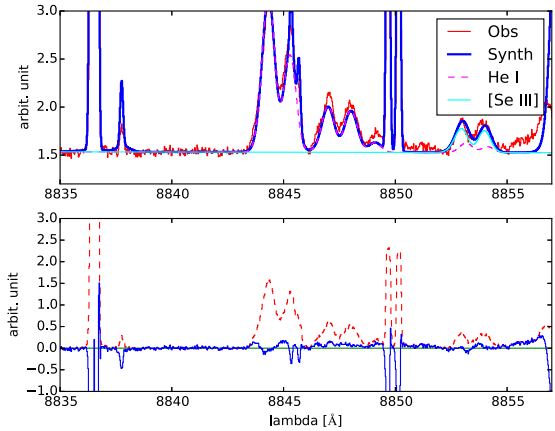}
\caption{PySSN fitting of the {\fseiii} $\lambda$8854.00 + {\hei} $\lambda$8854.11 lines. In the upper panel we show the observed spectrum (red line) and the different fits (magenta dashed line for {\hei}, cyan for {\fseiii}, and blue for the total spectrum); flux is in arbitrary units.  Note that a model of the telluric emission is included in the fit. In the lower panel, we show the residuals of the fit.}
\label{seiii_XSSN}
\end{center}
\end{figure}

Several authors reported that the identification of the {\fseiii} $\lambda$8854.00 line is quite difficult because of blending with a weak {\hei} $\lambda$8854.11 line which is at nearly the same wavelength \citep{pequignotbaluteau94, sharpeeetal07}. We used the PySSN spectral synthesis code to fit the nearby {\hei} lines and estimate the contribution of the {\hei} line to the blend. In Figure~\ref{seiii_XSSN} we show the best fit we have found. From this fit, a contribution of $\sim$75\% of the line was considered to come from the {\fseiii} emission. Another possible contaminant of this line is the {\fmniii} $\lambda$8854.2 line, but we discarded this line because we do not detect any other {\fmniii} line in the spectrum of NGC\,3918.

\subsubsection{Kr line identifications}

\begin{figure}
\begin{center}
\includegraphics[width=\columnwidth]{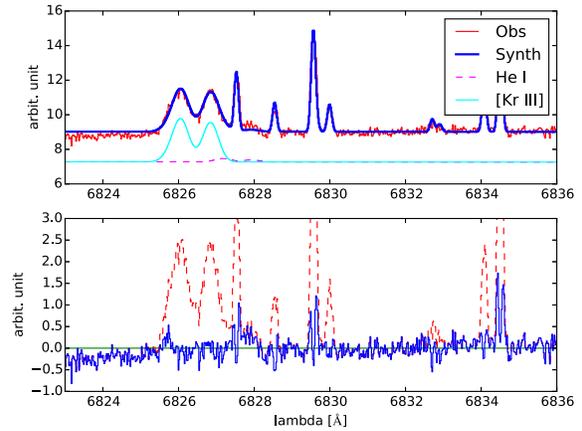}
\caption{PySSN fitting of the {\fkriii} $\lambda$6826.70 + {\hei} $\lambda$6827.88 lines. In the upper panel we show the observed spectrum (red line) and the different fits (magenta dashed line for {\hei}, cyan for {\fkriii}, and blue for the total spectrum); flux is in arbitrary units. Note that a model of the telluric emission is included in the fit. In the lower panel, we show the residuals of the fit.}
\label{kriii_XSSN}
\end{center}
\end{figure}

We identify lines of three ionization stages of Kr. \citet{pequignotbaluteau94} identify the {\fkriii} $\lambda$6826.70 line as a strong feature, partially blended with the faint {\ci} $\lambda$6828.12 and {\hei} $3s$ $^3S$--$16p$ $^3P^0$ $\lambda$6827.88 lines. Given the high ionization of NGC\,3918, we discarded the presence of {\ci} emission. We tentatively assume that the {\hei} line is contributing to the measured flux of the line, given the radial velocity measured. We detect other $3s$--$np$ transitions of {\hei} in our spectrum, but several of them are strongly affected by telluric absorptions and/or emissions and their intensities are unreliable. Therefore, we used the spectrum synthesis code PySSN to check how this line can affect the measured flux. In Fig.~\ref{kriii_XSSN} we can see that the {\hei} line is affecting the red end of the {\fkriii} emission; PySSN predicts that the {\hei} line should have a flux $\sim$20\% the flux of the {\fkriii} line, but, from a close inspection of Fig.~\ref{kriii_XSSN} only a portion of the {\hei} line ($\sim$25\% ) overlaps the {\fkriii} line and therefore, the contribution to the measured flux is within the uncertainties reported for the {\fkriii} line.  
\citet{sharpeeetal07} report that this line could also be potentially affected by {\ffeiv} $\lambda$6826.50. However, we do not detect other multiplet members that should be brighter than this line in our spectrum and therefore, this possibility was discarded. In Fig.~\ref{kriii_XSSN} we see that the telluric nightglow OH lines only slightly affect the measured flux of the {\fkriii} line and can be easily deblended. 
The {\fkriii} $\lambda$9902.30 line, which arises from the same level as  {\fkriii} $\lambda$6826.70, is also expected in the spectrum.
Indeed, a feature is found in our spectrum at the right wavelength, but the line is strongly blended with the {\cii} multiplet 17.02 $\lambda$9903.46 line. If we estimate the flux of {\fkriii} $\lambda$9902.30 by subtracting the flux of the {\cii} $\lambda$9903.46 line, which yields a C$^{2+}$ abundance consistent with what is obtained from the average of the other {\cii} recombination lines (see Sect.~\ref{orlsabund}), we  
still have a {\fkriii} $\lambda$9902.30 line too bright compared to the {\fkriii} $\lambda$6826.70 line (the theoretical line ratio {\fkriii} $\lambda$9902.30/$\lambda$6826.70 computed from state-of-the-art atomic data shown in Table~\ref{atomic_cels} is about ~0.08). This discrepancy might be explained in terms of either the occurrence of other lines contributing to the measured flux, or errors in the assumed {\cii} recombination coefficients. In either case, we cannot consider it as a robust identification and we then consider {\fkriii} $\lambda$6826.70  as the only confirmed line of the {\fkriii} spectrum.

The {\fkriv} $\lambda\lambda$ 5346.02 and 5867.74 nebular lines are usually the brightest neutron-capture lines in the optical spectra of PNe. As it can be seen in Fig.~\ref{selem_prof}, this is the case in NGC\,3918. In principle, these lines are well isolated and, at the excitation of NGC\,3918, are not affected by blends. On the other hand, the {\fkriv} $\lambda$ 6107.8 auroral line is extremely faint and it seems to be strongly blended with another line, which we identify as the brighest component of the {\niii} $5p$ $^2P_0$--$6d$ $^2D$ multiplet. The abundance computed from this line would be a factor of $\sim$4 larger than that derived from the nebular lines. The relative intensities of the {\fkriv} $\lambda\lambda$ 5346.02 and 5867.74 lines agree within 10\% with what is expected from their collision strengths. The excellent agreement between the observed and theoretical intensity ratios of the $\lambda\lambda$ 5346.02 and 5867.74 features strongly supports their identification with {\fkriv}.

We identify the {\fkrv} $\lambda$6256.1 line in our spectrum. \citet{pequignotbaluteau94} and \citet{zhangetal05} claim that this line can be affected by the {\cii} $\lambda$6257.18 line and/or the dielectronic {\cii} $\lambda$6256.52 line. Given our spectral resolution, the expected line profile and the radial velocities measured for other {\cii} lines, we can discard the presence of the {\cii} $\lambda$6257.18 line. Additionally, we discard the presence of the dielectronic {\cii} $\lambda$6256.52 line because we do not detect the brightest component of the multiplet at 6250.76 \AA. Unfortunately, we could not detect the {\fkrv} $\lambda$8243.39 line to compare the expected theoretical intensity ratio with {\fkrv} $\lambda$6256.1 line with the observed one because this line falls in a zone of the spectrum with a strong telluric absorption feature. Therefore, although the radial velocity is somewhat lower than expected, we consider the identfication of the {\fkrv} $\lambda$6256.1 line as quite robust. 

\subsubsection{Rb line identifications}

\begin{figure}
\begin{center}
\includegraphics[width=\columnwidth]{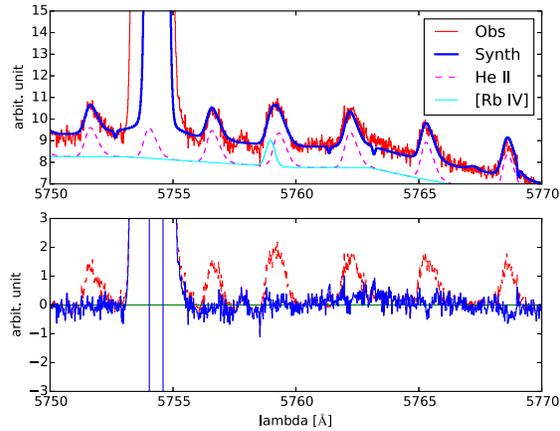}
\caption{PySSN fitting of the {\frbiv} $\lambda$5759.55 + {\heii} $\lambda$5759.74 lines. In the upper panel we show the observed spectrum (red line) and the different fits (magenta dashed line for {\heii}, cyan for {\frbiv}, and blue for the total spectrum). Lower panel is similar to that shown in Fig.~\ref{kriii_XSSN}.}
\label{rbiv_XSSN}
\end{center}
\end{figure}

We identify two lines of Rb in the spectrum of NGC\,3918. These identifications are complicated by blending with other features, especially the {\frbiv} $\lambda$5759.55 one. We reported the detection of this line as an excess intensity in the {\heii} 5--47 $\lambda$5759.74 line. We have assumed {\heii} $I$($\lambda$5759.55)/$I$({\hb})=0.0082 for $T_e$=12000 K and {\elecd}=1000 cm$^{-3}$ \citep{storeyhummer95} to correct for the contribution of this line. This give us a contribution of $\sim$36.4\% of the {\frbiv} line to the total measured flux. Moreover, we used  PySSN to check this by fitting the expected profile of a {\heii} line to this line, finding an excess of $\sim$35\% at the wavelength where we expect to find the {\frbiv} emission (see Fig.~\ref{rbiv_XSSN}). Therefore, we consider that this is a quite robust identification. Unfortunately, we could not detect the {\frbiv} $\lambda$9008.74 line, which arises from the same upper level as $\lambda$5759.74 line and that could give us an additional constraint for {\frbiv} line identification, because it falls in a zone with a strong telluric absorption band. 

Some authors report the identification of the {\frbv} $\lambda$5363.37 line in the spectra of PNe \citep{pequignotbaluteau94, sharpeeetal07}. The most likely contaminants to this line would be the [Ni~{\sc iv}] $\lambda$5363.35 and {\oii} $4f$ $F^2[4]^0_{7/2}$--$4d'$ $^2F_{7/2}$ $\lambda$5363.80 lines \citep{sharpeeetal07}. We do not detect other [Ni~{\sc iv}] lines belonging to the same multiplet and, therefore, we discard this line as contributing to the measured flux. On the other hand, regarding the possible detection of the {\oii} line, we inspected our spectra looking for other {\oii} lines arising from the lower level of the {\oii} $\lambda$5363.80 transition ($F^2[4]^0_{7/2}$) and tentatively detected a feature at $\sim$4609.15 \AA, corresponding to the $3d$ $^2D_{5/2}$--$4f$ $F^2[4]^0_{7/2}$ $\lambda$4609.4 transition, but its flux cannot be properly measured because it is strongly affected by a much brighter charge transfer feature. Other lines arising from the $F^2[4]^0_{7/2}$ level, such as {\oii} $\lambda\lambda$4282.02, 4312.11 or ending in this level, such as $\lambda\lambda$5321.97, 5361.74, 5363.80 or 5375.57 were not detected in our spectrum.  Additionally, the {\oii} line profiles are different to the profile observed and, therefore, the {\frbv} $\lambda$5363.37 identification is tentatively adopted. 

\subsubsection{Xe line identifications}

Several Xe identifications are considered probable in the spectrum of NGC\,3918. \citet{pequignotbaluteau94} identify {\fxeiii} $\lambda$5846.77 line as an excess intensity in the {\heii} 5--31 $\lambda$5846.66 line in NGC\,7027. Similarly, \citet{sharpeeetal07} also detect this blend in objects in their sample. Following the same procedure of the  {\frbiv} line identification and assuming the theoretical {\heii} $I$($\lambda$5846.66)/$I$({\hb})=0.0247 ratio for $T_e$=12000 K and {\elecd}=1000 cm$^{-3}$ \citep{storeyhummer95} to correct for the contribution of this line, we find a contribution of $\sim$8\% of the {\fxeiii} line to the total measured flux. As this is a rather uncertain estimation, and it lies within the uncertainties of the measured flux, we used the emission line spectral synthesis code PySSN to fit the expected profile of an {\heii} line finding that, in principle, there is no substantial contribution of the {\fxeiii} line to the measured flux. Therefore, we cannot conclude that {\fxeiii} $\lambda$5846.77 is detected in our spectrum. Nevertheless, we do the exercise of computing an upper limit to the Xe$^{2+}$ abundance adopting the flux excess measured in the {\heii} line as the flux of the {\fxeiii} line.

We identify two possible {\fxeiv} lines at $\lambda\lambda$5709.10, 7535.4 \AA. The {\fxeiv} $\lambda$5709.21 is well separated from the widely observed {\nii} multiplet 3 $\lambda$5710.77 line. Other potential contaminants, such as [Fe~{\sc i}] $\lambda$5708.91 and the intercombination Fe~{\sc ii}] $\lambda$5709.04 line, can be discarded since no other lines of these ions are observed in the spectrum of NGC\,3918 (a situation similar to what \citet{sharpeeetal07} report for most of their objects). \citet{sharpeeetal07} argue that all the lines most likely to affect the {\fxeiv} $\lambda$7535.4 emission come from low ionization ions (Fe~{\sc ii}] $\lambda$7535.4, {\neii} $\lambda$7534.75, {\fcrii} $\lambda$7534.80, {\nii} $\lambda$7535.10 and Ne~{\sc i} $\lambda$7535.77). Given the profile of the line we detect and the excitation of NGC\,3918, we can safely discard all of them as possible contributors to the measured flux and, hence, we consider {\fxeiv} $\lambda$7535.4 as a robust identification. The excellent agreement between the observed and theoretical intensities of the $\lambda\lambda$ 5709.10 and 7535.4 features strongly supports their identification with {\fxeiv}. The intensity ratio of these lines agree within 17\% with what is expected from their collisional strengths and is consistent with the uncertainties of these lines.

We do not detect any [Xe~{\sc v}] line in our spectra. \citet{sharpeeetal07} report the possible detection of the [Xe~{\sc v}] $\lambda$7076.8 line in the two most excited PNe of their sample. Although we do detect an emission feature at $\sim$7076.5 \AA, the feature is strongly affected by telluric absorptions, making its flux unreliable. On the other hand, we do not detect any line near 5228.8 \AA\ or near 6998.7 \AA, which correspond to the laboratory wavelengths of the [Xe~{\sc v}] $^3P_{1,2}$--$^1D_2$ transitions. In the first case, no emission was detected and in the second case, telluric absorption lines preclude the identification of any emission line. 

\citet{pequignotbaluteau94} and \citet{sharpeeetal07} report the detection of {\fxevi} $\lambda$6408.89 line in the spectrum of NGC\,7027 on the red wing of the {\heii} 5--15 $\lambda$6406.38 line. We also detect two extremely faint features at a similar position. We discard the identification of these lines in terms of a double-peaked line of {\ffeiii} $\lambda$6408.50, owing to the low number of {\ffeiii} identifications in our spectrum and to the different line profile of {\ffeiii} lines. Alternative identifications could be the {\civ} $\lambda\lambda$6408.1, 6408.8 lines.  \citet{sharpeeetal07} discard these identifications owing to the high excitation of these lines, although they report the detection of other {\civ} lines in the spectrum of NGC\,7027. Since we detect other brighter lines of the $9g-{\rm n}h$ series, we confirm the identification of the {\civ} $9g$ $^2G$--$17h$ $^2H_0$ $\lambda$6408.1. However, the second line can hardly be {\civ} $9h$ $^2H_0$--$17i$ $^2I$ $\lambda$6408.8, since there are no other lines of the $9h-{\rm n}i$ series in our spectra. The identification of this line as {\fxevi} $\lambda$6408.90 is also highly questionable, given the faintness of the line; however, when we compute the Xe$^{5+}$ abundance from this line, we find a reasonable result (see Sect.~\ref{ab_cels}).  

Since the intensities of several {\emph n}-capture element lines are corrected by deblending other nebular or telluric lines, in Table~\ref{new_inten} we show the detected {\emph n}-capture element lines with line intensities corrected for contamination from other features.
In the following, we comment on identifications of individual neutron-capture elements.

\setcounter{table}{3}
\begin{table}
\centering
\caption{Corrected line ratios (I(H$\beta$) = 100) for {\emph n}-capture element lines in NGC\,3918.}
\label{new_inten}
\begin{tabular}{lccc}
\hline
$\lambda_0$ (\AA\ ) & Ion  & $I(\lambda)/I({\rm H}\beta)$ & Error(\%) \\
\hline
5346.02 & {\fkriv} & 0.105 & 6 \\
5363.37 & {\frbv} & 0.0045 & 38 \\
5709.21 & {\fxeiv} & 0.0065 & 27 \\
5759.55 & {\frbiv} & 0.0047 & : \\
5846.77 & {\fxeiii} & 0.0021 & : \\
5867.74 & {\fkriv}  & 0.131 & 15 \\
6256.10 & {\fkrv}  & 0.0078 & 23 \\
6408.80 & {\fxevi} & 0.0039 & : \\
6826.70 & {\fkriii} & 0.0284 & 12 \\
7535.40 & {\fxeiv} & 0.0076 & 23 \\
8845.00 & {\fseiii} & 0.0058 & :\\
\hline
\end{tabular}
\end{table}

\section{Physical conditions}\label{physcond}

The large number of emission lines identified and measured in the spectrum of NGC\,3918 allows us to derive physical conditions using multiple emission-line ratios. 
The computations of physical conditions were carried out with PyNeb $v$1.0.9 \citep{luridianaetal15}, a python-based package
dedicated to the analysis of emission line spectra. The methodology followed for the derivation of the electron density, {\elecd}, and the electron temperature, {\te}, has been described in previous papers of our group. We have updated the atomic data set to state-of-the-art atomic data, presented in Tables~\ref{atomic_cels} and~\ref{atomic_rls} for collisionally excited lines (CELs) and optical recombination lines (ORLs), respectively. Errors in the diagnostics were computed via Monte Carlo simulations.  We generate 1500 random values for each line intensity using a Gaussian distribution centered in the observed line intensity with a sigma equal to the associated uncertainty. For higher number of Monte Carlo simulations, the errors in the computed quantities remain constant.
The electron temperatures and densities are presented in Table~\ref{phy_cond}.

For the calculation of ionic abundances (see Section~\ref{chemabun}) we assumed a three-zone ionization scheme. For {\elecd}, we obtained very similar results from the different diagnostics; therefore, we adopted the average of {\elecd}({\foii}), {\elecd}({\fsii}), {\elecd}({\fcliii}) and {\elecd}({\fariv}) as representative of the whole nebula. We adopted the average of electron temperatures obtained from {\fnii}, {\fsii} and {\foii} lines as representative of the low-ionization zone (IP $<$ 17 eV); we refer to this value as {\te}(low).
Similarly, the average of electron temperatures obtained from {\foiii}, {\fariii}, {\fsiii} and {\fcliv} lines was assumed as representative of the medium-ionization zone (17 eV $<$ IP $<$ 39 eV) and designated {\te}(mid). Finally, the electron temperature from {\farv} (see below) was adopted as representative of the high-ionization zone (IP $>$ 39 eV) and designated {\te}(high) (see Table~\ref{phy_cond}).

Following Eqs. (1) and (2) by \citet{liuetal00}, we corrected the intensity of the auroral {\fnii} $\lambda$5755 line and the trans-auroral {\foii} $\lambda$$\lambda$7320+30 lines from the recombination contribution. We assumed preliminary computations of the N$^{++}$/H$^+$ and O$^{++}$/H$^+$ ratios obtained from {\nii} multiplets 3, 39 and 48 and from the {\oii} multiplet 1 to compute the recombination contributions to {\fnii} $\lambda$5755 line and {\foii} $\lambda$$\lambda$7320+30 lines, respectively. These contributions led to a $\sim$6\% correction in the {\fnii} line and in the {\foii} lines, which translate in temperatures lower by $\sim$300 K in both temperature diagnostics. Additionally, we also corrected the contribution of the {\heii} 5--49 $\lambda$5754.67 line ($I$({\heii} 5.49)/$I$({\hb})$\sim$0.0078). Given the high ionization degree of NGC\,3918, the recombination contribution to the auroral $\lambda$4363 line may also be non-negligible; we estimated it using Eq. (3) of \citet{liuetal00}, where the O$^{3+}$/H$^+$ ratio was estimated assuming O$^{3+}$/H$^+$=(He/He$^+$)$^{2/3}$$\times$(O$^+$/H$^+$+O$^{++}$/H$^+$). Using the values derived for He$^+$/H$^+$ and He$^{++}$/H$^+$ from ORLs (see Table~\ref{he_ab}) and O$^+$/H$^+$ and O$^{++}$ /H$^+$ from CELs (see Table~\ref{ionic}), this contribution amounts to $\sim$1\%, which has almost no effect on the determination of {\te}({\foiii}) and therefore, was not considered.

\setcounter{table}{4}
\begin{table}
\centering
\caption{Atomic data set used for collisionally excited lines.}
\label{atomic_cels}
\begin{tabular}{lcc}
\hline
& Transition  & Collisional \\
Ion & Probabilities & Strengths \\
\hline
N$^+$ & \citet{froesefischertachiev04} & \citet{tayal11} \\
O$^+$ & \citet{froesefischertachiev04} & \citet{kisieliusetal09} \\
O$^{2+}$ &  \citet{froesefischertachiev04} &  \citet{storeyetal14} \\
                &  \citet{storeyzeippen00} &  \\
Ne$^{2+}$ & \citet{galavisetal97} & \citet{mclaughlinbell00} \\
Ne$^{3+}$ & \citet{butlerzeippen89} & \citet{giles81} \\
          &  \citet{bhatiakastner88} & \\
Ne$^{4+}$ & \citet{galavisetal97} & \citet{danceetal13} \\
          &  \citet{bhatiadoschek93} & \\
          &  Unknown & \\
Na$^{3+}$ & \citet{beckeretal89} & \citet{giles81} \\
         &  \cite{bhatiakastner88} & \\
S$^+$ & \citet{podobedovaetal09} & \citet{tayalzatsarinny10} \\
S$^{2+}$ &  \citet{podobedovaetal09} & \citet{tayalgupta99} \\
Cl$^{+}$ & \citet{mendozazeippen83} & \citet{tayal04} \\
Cl$^{2+}$ & \citet{mendoza83} & \citet{butlerzeippen89} \\
Cl$^{3+}$ & \citet{kaufmansugar86} & \citet{galavisetal95} \\
         &  \cite{mendozazeippen82b} & \\
         &  \cite{ellismartinson84} & \\
Ar$^{2+}$ & \citet{mendoza83} & \citet{galavisetal95} \\
          &  \citet{kaufmansugar86} & \\
Ar$^{3+}$ & \citet{mendozazeippen82a} & \citet{ramsbottombell97} \\
Ar$^{4+}$ & \citet{mendozazeippen82b} & \citet{galavisetal95} \\
          &  \citet{kaufmansugar86} & \\
          &  \citet{lajohnluke93} & \\
K$^{3+}$ & \citet{kaufmansugar86} & \citet{galavisetal95} \\
          &  \citet{mendoza83} & \\
K$^{4+}$ & \citet{kaufmansugar86} & \citet{butleretal88} \\
          &  \citet{mendoza83} & \\
Ca$^{4+}$ & \citet{kaufmansugar86} & \citet{galavisetal95} \\
          &  \citet{mendoza83} & \\
Fe$^{2+}$ & \citet{quinet96} & \citet{zhang96} \\
          &  \citet{johanssonetal00} & \\
Fe$^{3+}$ & \citet{froesefischeretal08} & \citet{zhangpradhan97} \\
Fe$^{4+}$ & \citet{naharetal00} & \citet{ballanceetal07} \\
Fe$^{5+}$ & \citet{chenpradhan00} & \citet{chenpradhan99} \\
Fe$^{6+}$ & \citet{witthoeftbadnell08} & \citet{witthoeftbadnell08} \\
Se$^{2+}$ & \citet{biemonthansen86a} & \citet{schoning97} \\
Kr$^{2+}$ & \citet{biemonthansen86a} & \citet{schoning97} \\
Kr$^{3+}$ & \citet{biemonthansen86b} & \citet{schoning97} \\
Kr$^{4+}$ & \citet{biemonthansen86b} & \citet{schoning97} \\
Rb$^{3+}$ & \citet{biemonthansen86b} & \citet{schoning97}$^{\rm a}$ \\
Rb$^{4+}$ & \citet{biemonthansen86a} & \citet{schoning97}$^{\rm b}$ \\
Xe$^{2+}$ & \citet{biemontetal95} & \citet{schoningbutler98} \\
Xe$^{3+}$ & \citet{biemontetal95} & \citet{schoningbutler98} \\
Xe$^{5+}$ & \citet{biemontetal95} & \citet{schoningbutler98} \\
\hline
\end{tabular}
\begin{description}
\item[$^{\rm a}$] Scaled from Kr$^{3+}$ effective collision strengths.
\item[$^{\rm b}$] Scaled from Kr$^{4+}$ effective collision strengths.
\end{description}
\end{table}

\setcounter{table}{5}
\begin{table}
\centering
\begin{minipage}{180mm}
\caption{Atomic data set used for recombination lines.}
\label{atomic_rls}
\begin{tabular}{lc}
\hline
Ion & Recombination Coefficients \\
\hline
H$^{+}$ &  \citet{storeyhummer95}  \\
He$^{+}$ &  \citet{porteretal12, porteretal13}  \\
He$^{2+}$ &  \citet{storeyhummer95}  \\
C$^{2+}$ & \citet{daveyetal00}  \\
N$^{2+}$ & \citet{fangetal11, fangetal13} \\´
		&\citet{escalantevictor92}  \\
		&\citet{kisieliusstorey02}  \\
O$^{2+}$ &  \citet{storey94}  \\
                &  \citet{liuetal95}  \\
CNO$^{3+}$ & \citet{pequignotetal91}  \\
                & \citet{nussbaumerstorey84}  \\
CO$^{4+}$ & \citet{pequignotetal91}  \\
N$^{4+}$ & \citet{pequignotetal91}  \\
                & \citet{nussbaumerstorey84}  \\
Ne$^{2+}$ & \citet{kisieliusetal98}  \\
	         & Kisielius \& Storey (private communication)  \\
\hline
\end{tabular}
\end{minipage}
\end{table}

\setcounter{table}{6}
\begin{table}
\caption{Physical conditions.}
\label{phy_cond}
\begin{tabular}{lc}
\noalign{\smallskip} \noalign{\smallskip} \noalign{\hrule} \noalign{\smallskip}
Diagnostic 	&   \\
\noalign{\smallskip} \noalign{\hrule} \noalign{\smallskip}
{\elecd} (cm$^{-3}$) 	&   \\
\noalign{\smallskip} \noalign{\hrule} \noalign{\smallskip}
{\fsii} $\lambda$6716/$\lambda$6731	& 5000$^{+3000}_{-2100}$	\\[3pt]
{\foii} $\lambda$3726/$\lambda$3729   	& 5600$^{+2700}_{-1800}$	\\[3pt]
{\fcliii} $\lambda$5517/$\lambda$5537   & 6000$^{+1100}_{-1000}$	\\[3pt]
{\fariv} $\lambda$4711/$\lambda$4741  	& 6500$^{+1300}_{-1200}$	\\[3pt]
{\bf Adopted}			&	{\bf 6200$\pm$1250	}		\\[3pt]
\noalign{\smallskip} \noalign{\hrule} \noalign{\smallskip}
{\te} (K) 	&   \\
\noalign{\smallskip} \noalign{\hrule} \noalign{\smallskip}
{\fnii} $\lambda$5755/$\lambda$6548 & 10950$\pm$700	\\[3pt]
{\foii} $\lambda$$\lambda$3726+29/$\lambda$$\lambda$7320+30 &	11350$\pm$2800		\\[3pt]
{\fsii} $\lambda$$\lambda$4068+76/$\lambda$$\lambda$6716+31 & 10650$\pm$1800	\\[3pt]
{\bf $T_e$(low) (adopted)} & {\bf 11000$\pm$1350} \\[3pt]
{\foiii} $\lambda$4363/$\lambda$4959 & 12800$\pm$400 	\\[3pt]
{\fsiii} $\lambda$6312/$\lambda$9069 & 12550$\pm$950	\\[3pt]
{\fariii} $\lambda$5192/$\lambda$7136 & 11200$\pm$400	\\[3pt]
{\fcliv} $\lambda$5323/$\lambda$$\lambda$7531+8046 & 11950$\pm$450	\\[3pt]
{\bf $T_e$(mid) (adopted)} & {\bf 12100$\pm$300} \\[3pt]
{\fariv} $\lambda$$\lambda$7170+263/$\lambda$$\lambda$4711+40 & 17900$\pm$1250	\\[3pt]
{\farv} $\lambda$4626/$\lambda$$\lambda$6425+7005 & 15400$\pm$800	\\[3pt]
{\bf $T_e$(high) (adopted)} & {\bf 15400$\pm$800} \\[3pt]
$T_e$(BJ) & \bf 11750$\pm$1400 \\[3pt]
\noalign{\smallskip} \noalign{\hrule} \noalign{\smallskip}
\end{tabular}
\end{table}

\begin{figure}
\begin{center}
\includegraphics[width=\columnwidth]{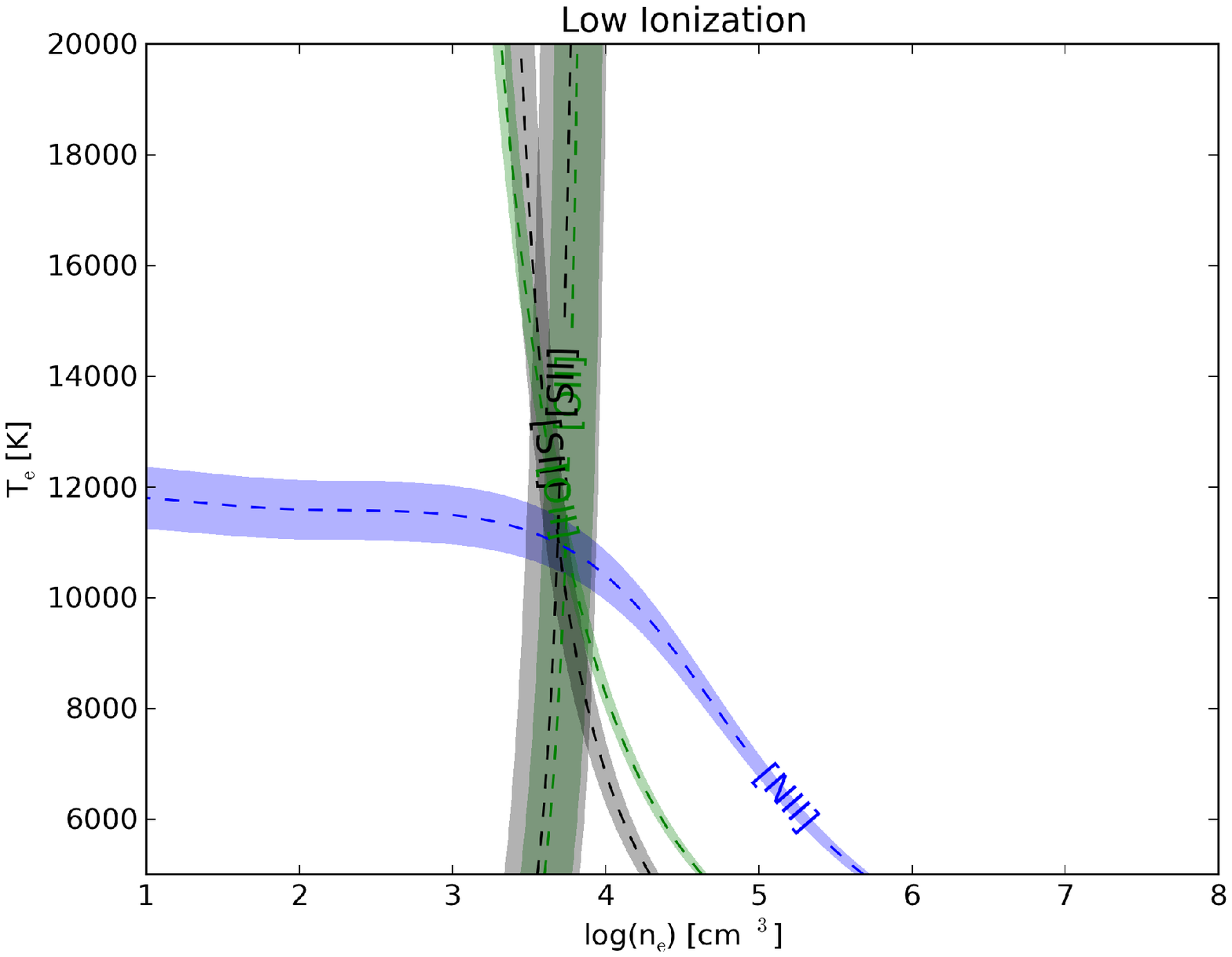}
\includegraphics[width=\columnwidth]{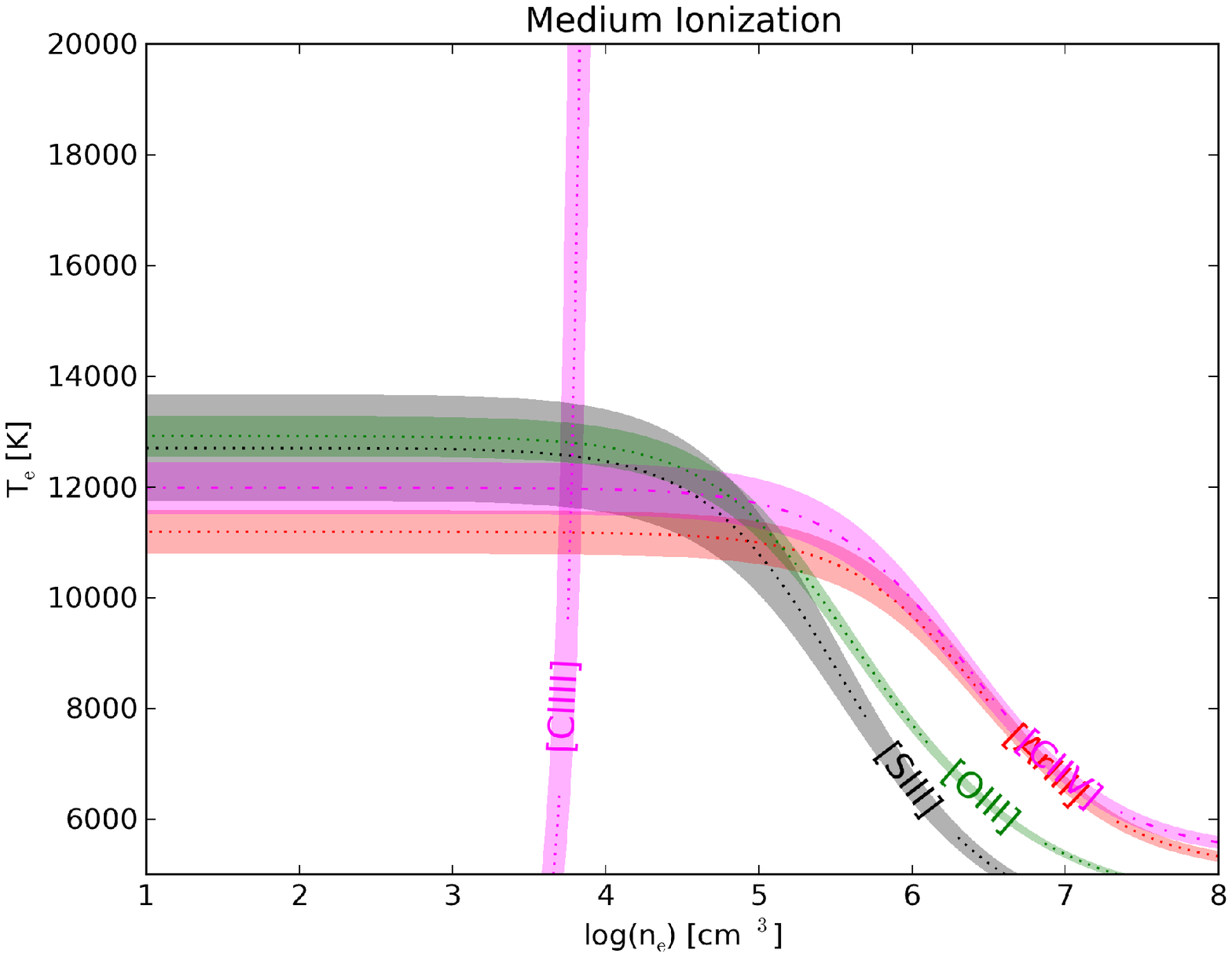}
\includegraphics[width=\columnwidth]{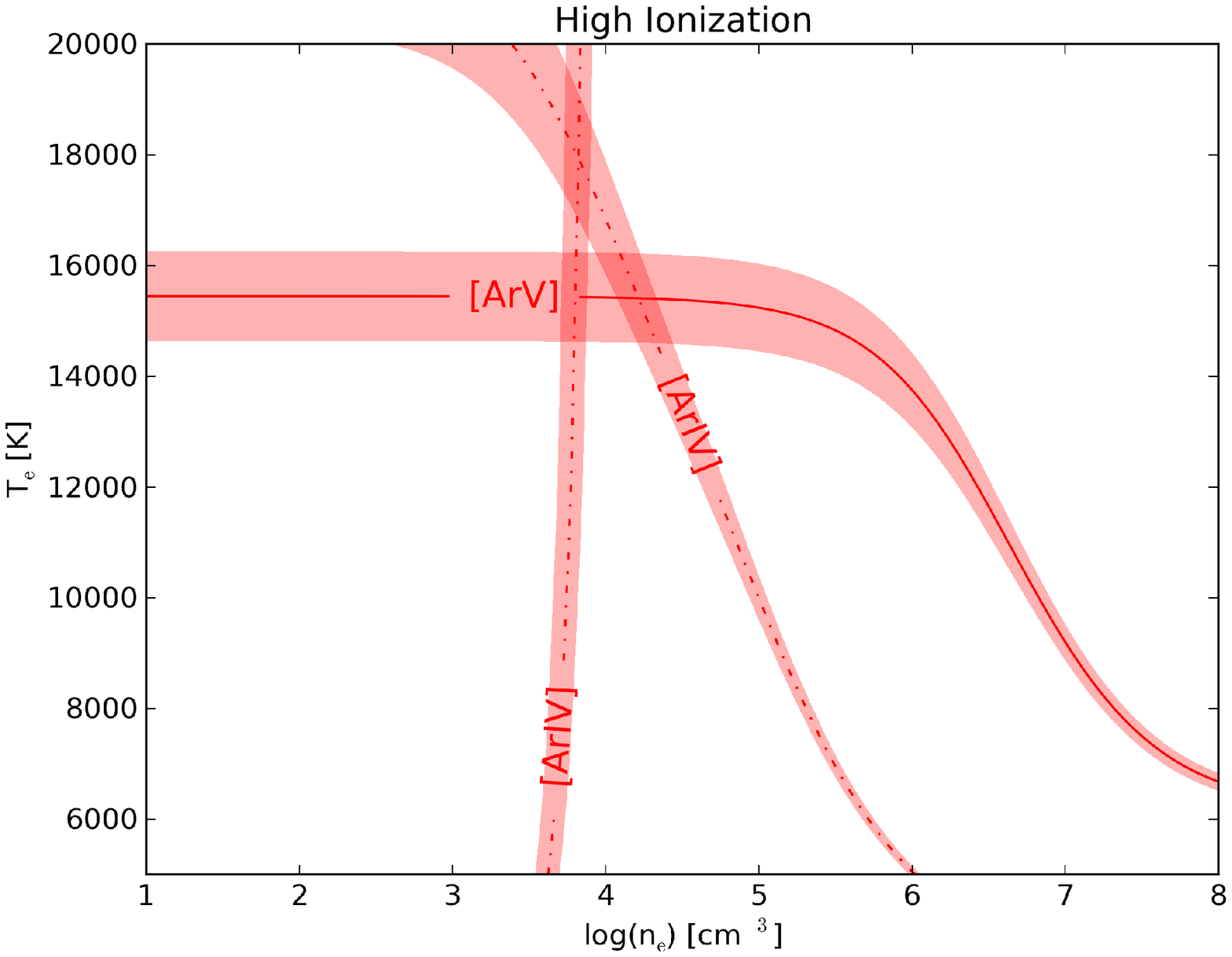}
\caption{Diagnostic diagrams for the low (upper panel), medium (middle panel) 
and high (lower panel) ionization zones in NGC\,3918.}
\label{diags}
\end{center}
\end{figure}

In Figure~\ref{diags} we show the diagnostics diagrams obtained for each ionization zone in NGC\,3918. It is worth noting the agreement between the different diagnostics except for the high-ionization zone, where {\fariv} temperature diagnostic gives a very high {\te}. We discarded {\te}({\fariv}) because we notice in a previous paper \citep[see section 4.3 of][]{garciarojasetal13} that the ratio between
collisional strengths for {\fariv} $\lambda$$\lambda$7170+263 and $\lambda$$\lambda$4711+40 lines might be unreliable. Therefore, only the {\farv} diagnostic was considered for the high-ionization zone.

The electron temperature can also be derived from the ratio between the Balmer discontinuity end the {\hi} lines belonging to the Balmer serie. We use the expression given by \citet{liuetal01} to compute the electron temperature {\te}(BJ) from the ratio of the Balmer discontinuity to $I$(H11).

The physical conditions reported in Table~\ref{phy_cond} agree within the uncertainties with the physical conditions reported by \citet{cleggetal87} and \citet{tsamisetal03b}.   

\section{Chemical abundances}\label{chemabun}

\subsection{Ionic abundances from CELs}\label{ab_cels}

Ionic abundances from CELs were computed using PyNeb \citep{luridianaetal15} and the atomic data in Table~\ref{atomic_cels}. To perform the computations required for this paper, we updated PyNeb by including several new ions; the updated version of PyNeb ($v$1.0.9) was uploaded to the PyNeb website\footnote{http://www.iac.es/proyecto/PyNeb/}. For highly ionized Fe ions, we adopt the data recommended by \citet{delgadoingladarodriguez14}, and for neutron-capture elements we use the atomic data references compiled by \citet{sharpeeetal07}. Errors in the line fluxes and the physical conditions were propagated via Monte Carlo simulations. As pointed out in Sect.~\ref{physcond}, we use a three-zone scheme of the nebula, adopting a unique {\elecd} value in the three zones and {\te}(low) for ions with IP$<$17 eV (i.~e. N$^+$, O$^+$, S$^+$ and Fe$^{2+}$), {\te}(mid) for ions with 17 eV $<$ IP $<$ 39 eV (i.~e. O$^{2+}$, S$^{2+}$, Cl$^{2+}$, Cl$^{3+}$, Ar$^{2+}$, Fe$^{3+}$, Se$^{2+}$, Kr$^{2+}$, Kr$^{3+}$, Xe$^{2+}$ and Xe$^{3+}$), and {\te}(high) for ions with IP $>$ 39 eV (i.~e. Ne$^{2+}$, Ne$^{3+}$, Ne$^{4+}$, Na$^{3+}$, Ar$^{3+}$, Ar$^{4+}$, K$^{3+}$, K$^{4+}$, Ca$^{4+}$, Fe$^{4+}$, Fe$^{5+}$, Fe$^{6+}$, Kr$^{4+}$, Rb$^{3+}$, Rb$^{4+}$ and Xe$^{5+}$). Ionic abundances are presented in Table~\ref{ionic}, along with the lines used to compute them. We avoid using auroral lines in abundance computations because they do not affect significantly the computed abundances but can introduce some scatter.

Since transition probabilities and effective collision strengths for several of the detected Fe-peak species ({\fcrii}, {\fcriii}, {\fmniv}, {\fmnv}, {\fcoiv}, {\fniqv}, {\fcuv}, and {\fcuvi}) are unknown, we are unable to determine the ionic or elemental abundances of these species. Most of these ions require large model atoms (30-50 levels), as used for Fe ions \citep[e.~g.][]{rodriguezrubin05, delgadoingladarodriguez14} in order to accurately determine the fractional populations of the upper levels of the observed transitions. Gas-phase abundances of these species would provide valuable information regarding the chemical content of dust grains in PNe and AGB star winds. New atomic data determinations are needed to make such investigations possible.

\setcounter{table}{7}
\begin{table}
\begin{scriptsize}
\caption{CELs ionic abundances.}
\label{ionic}
\begin{tabular}{llc}
\noalign{\smallskip} \noalign{\smallskip} \noalign{\hrule} \noalign{\smallskip}
Ion 	& Lines used & 12 + log(X$^{i+}$/H$^+$)$^{\rm a}$ \\
\noalign{\smallskip} \noalign{\hrule} \noalign{\smallskip}
N$^{+}$	   	& {\fnii} $\lambda$$\lambda$6548+84							&  7.12$^{+0.16}_{-0.10}$ 	\\[3pt]
O$^{+}$	   	& {\foii} $\lambda$$\lambda$3726+29							&  7.58$^{+0.29}_{-0.17}$ 	\\[3pt]
O$^{2+}$   	& {\foiii} $\lambda$$\lambda$4959, 5007						&  8.47$\pm$0.04        	\\[3pt]	
Ne$^{2+}$  	& {\fneiii} $\lambda$$\lambda$3868, 3967						&  7.80$\pm$0.04		\\[3pt]
Ne$^{3+}$  	& {\fneiv} $\lambda$$\lambda$4714+15, 4724+25				&  7.50$\pm$0.15 		\\[3pt]
Ne$^{4+}$  	& {\fnev} $\lambda$$\lambda$3345, 3425 						&  7.20$\pm$0.08 		\\[3pt]
Na$^{3+}$  	& {\fnaiv} $\lambda$$\lambda$3242, 3362						&  5.53$\pm$0.08 		\\[3pt]
S$^{+}$	   	& {\fsii} $\lambda$$\lambda$6717+31							&  5.57$^{+0.20}_{-0.12}$    \\[3pt]
S$^{2+}$   		& {\fsiii} $\lambda$$\lambda$8829, 9069						&  6.45$\pm$0.07		\\[3pt]
Cl$^{+}$   		& {\fclii} $\lambda$ 9123									&  3.89$\pm$0.11   		\\[3pt]
Cl$^{2+}$  		& {\fcliii} $\lambda$$\lambda$3353, 5517+37, 8434				&  4.84$\pm$0.05		\\[3pt]
Cl$^{3+}$  		& {\fcliv} $\lambda$$\lambda$7531, 8045						&  4.87$\pm$0.04 		\\[3pt]
Ar$^{2+}$  	& {\fariii} $\lambda$$\lambda$3109, 7136, 7751					&  6.10$\pm$0.04		\\[3pt]
Ar$^{3+}$  	& {\fariv} $\lambda$$\lambda$4711+40						&  6.10$\pm$0.04		\\[3pt]
Ar$^{4+}$  	& {\farv} $\lambda$$\lambda$6435, 7005	 					&  5.21$\pm$0.06	 	\\[3pt]
K$^{3+}$   		& {\fkiv} $\lambda$$\lambda$6102, 6796						&  4.14$\pm$0.06		\\[3pt]
K$^{4+}$   		& {\fkv} $\lambda$$\lambda$4123+63							&  4.10$\pm$0.07     		\\[3pt]
Ca$^{4+}$   	& {\fcav} $\lambda$5309									&  3.76$\pm$0.06      		\\[3pt]
Fe$^{2+}$  	& {\ffeiii} $\lambda$$\lambda$4701+55, 4881, 5270			         &  4.01$^{+0.18}_{-0.14}$   \\[3pt]
Fe$^{3+}$  	& {\ffeiv}  $\lambda$6740									&  4.35:				\\[3pt]
Fe$^{4+}$  	&{\ffev }  $\lambda$$\lambda$3891+95, 4181, 4227				&  4.38$\pm$0.07		\\[3pt]
Fe$^{5+}$  	&{\ffevi }  $\lambda$$\lambda$5278, 5335+70, 5424+27+85, 5631+77	&  4.25$\pm$0.06  		\\[3pt]
Fe$^{6+}$  	&{\ffevii }  $\lambda$$\lambda$5158, 5276, 5720. 6086				&  4.14$\pm$0.06  		\\[3pt]
Se$^{2+}$  	& {\fseiii} $\lambda$8854									&  2.79$^{+0.15}_{-0.23}$   	\\[3pt]
Kr$^{2+}$  		& {\fkriii} $\lambda$6827									&  3.23$\pm$0.06		\\[3pt]
Kr$^{3+}$  		& {\fkriv} $\lambda$$\lambda$5346, 5868						&  3.62$\pm$0.05		\\[3pt]
Kr$^{4+}$  		& {\fkrv} $\lambda$6256									&  2.60$\pm$0.11	     	\\[3pt]
Rb$^{3+}$  	& {\frbiv} $\lambda$5760									&  2.41:	\\[3pt]
Rb$^{4+}$  	& {\frbv} $\lambda$5364									&  1.83$^{+0.17}_{-0.28}$    \\[3pt]
Xe$^{2+}$  	& {\fxeiii} $\lambda$5847									&  $<$1.47		\\[3pt]
Xe$^{3+}$  	& {\fxeiv} $\lambda$$\lambda$5709, 7535						&  2.41$\pm$0.08	\\[3pt]
Xe$^{5+}$  	& {\fxevi} $\lambda$6409									&  1.37:    \\[3pt]
\noalign{\smallskip} \noalign{\hrule} \noalign{\smallskip}
\end{tabular}
\begin{description}
\item[$^{\rm a}$] Only lines with intensity uncertainties lower than 40\% have been considered, except in the case of Fe$^{3+}$ and neutron-capture lines, where we used all the available lines (see text).
\end{description}
\end{scriptsize}
\end{table}

\subsection{Ionic abundances from recombination lines}\label{ab_rls}

In the following subsections, we present ionic abundances for He, C, N, O and Ne derived from ORLs. We adopt the physical conditions derived from CELs to compute the abundances, as ORLs depend only weakly on electron temperature and are essentially independent on the electron density under the low density conditions of NGC\,3918. 

\subsubsection{He$^{+}$ and He$^{++}$ abundances}\label{heabund}

We detect several {\hei} emission lines in the spectrum of
NGC\,3918. These lines arise mainly from recombination, but
some of them can be affected by collisional excitation and self-absorption
effects.
We use the effective recombination coefficients compiled in Table~\ref{atomic_rls} for {\hi} and {\hei}. 
Both collisional contribution effects and the optical depth in the triplet lines are included in the computations. 
Using PyNeb, we determine the He$^+$/H$^+$ ratio from the three brightest {\hei} emission lines: $\lambda$$\lambda$4471, 5876 and 6678.

We measure multiple {\heii} emission lines in the spectrum of NGC\,3918, but, to compute the He$^{2+}$/H$^+$ ratio, we use only the 11 brightest ones among those not affected by telluric features or line blending 
($\lambda$$\lambda$3203, 4100, 4200, 4339, 4542, 4686, 4859, 5412, 6560, 7593 and 8237 \AA).  
The computation adopts the recombination coefficients computed by \citet{storeyhummer95}. 
The adopted He$^+$/H$^+$ and He$^{2+}$/H$^+$ ratios are presented in Table~\ref{he_ab}.

\setcounter{table}{8}
\begin{table}
\caption{He$^+$ and He$^{2+}$ abundances.}
\label{he_ab}
\begin{tabular}{lc}
\noalign{\smallskip} \noalign{\hrule} \noalign{\smallskip}
Ion			& 12+log(He$^{i+}$/H$^+$)\\
\noalign{\smallskip} \noalign{\hrule} \noalign{\smallskip}
He$^+$/H$^+$		& 10.77$\pm$0.02    	 	\\ 
He$^{2+}$/H$^+$		& 10.63$\pm$0.01    	\\ 
\noalign{\smallskip} \noalign{\hrule} \noalign{\smallskip}
\end{tabular}
\end{table}

\subsubsection{Ionic abundances from C, N, O and Ne recombination lines}\label{orlsabund}

A well known problem of nebular astrophysics is the difference between the chemical composition of photoionized plasmas as derived from collisionally excited lines and from recombination lines (ORLs). This discrepancy is parametrized in terms of an abundance discrepancy factor (ADF), which is defined as:

\begin{equation}
{\rm ADF}(X^{i+}) = (X^{i+}/H^+)_{\rm ORLs} / (X^{i+}/H^+)_{\rm CELs}, 
\end{equation}

\noindent where X$^{i+}$/H$^+$ is the abundance of $i$-times ionized element X relative to H$^+$. ADFs may reach values of 2-3 in {\hii} regions \citep{garciarojasesteban07} and up to 120 in PNe, with an average value of about 3 \citep{mcnabbetal13, corradietal15}.

Thanks to the high quality of our data, we could measure a large number of permitted lines of heavy-element ions such as {\oi}, {\oii}, {\oiii}, {\oiv}, {\cii}, {\ciii}, {\civ}, {\nii}, {\niii}, {\niv}, {\neii}, {\silii}, {\siliii}, {\mgi}, and {\mgii},
many of them detected for the first time in NGC\,3918, and compute the ADF for several ions. Unfortunately, many of these permitted lines are affected by fluorescence or blended with telluric emission lines, making their measured intensities unreliable. A discussion on the mechanisms of formation of permitted lines can be found in 
\citet[][and references therein]{estebanetal98, estebanetal04}. We could not estimate the Mg$^{2+}$/H$^+$ ratio because the pure RL of {\mgii} $\lambda$4481 is affected by charge transfer in the CCD and its flux is unreliable. 

The abundances are derived using the atomic data compiled in Table~\ref{atomic_rls}. For C$^{3+}$, N$^{3+}$, N$^{4+}$, and O$^{3+}$ we take also into account the contribution of dielectronic recombination computed by \citet{nussbaumerstorey84}. We selected the lines formed by pure recombination to compute ionic abundances from ORLs in PNe \citep[see][]{liuetal00, tsamisetal04}.  
For {\civ}, {\niii} and {\niv} lines and some multiplets of {\oiii} and {\oiv}, we adopt 
the log($gf$)s given by the computations in the Atomic Line List 
v2.05b18\footnote{webpage at: http://www.pa.uky.edu/$\sim$peter/newpage/}. The log($gf$) values necessary to compute abundances from the remaining lines for which we have assumed LS-coupling have been adopted from the local thermodynamic equilibrium (LTE) computation predictions from \citet{wieseetal96}.

\paragraph{C ORLs.}

Owing to the high ionization of NGC\,3918, we detect lines of {\cii}, {\ciii} and {\civ} that are excited by pure recombination and therefore are suitable for abundance determinations \citep[see][and references therein]{garciarojasetal09}. 
In Table~\ref{C_rls} we show the ionic abundances from recombination lines of C. The results obtained for the different multiplets of each given ion clearly show the excellent agreement between the different lines, strengthening the hypothesis of the recombination origin of these lines. \citet{tsamisetal04} computed the ADF for C$^{2+}$ and C$^{3+}$ by comparing the abundances obtained from {\cii} and {\ciii} ORLs with the abundances obtained from {\it IUE} UV data of the C~{\sc iii}] $\lambda\lambda$1906+09 CELs and the {\civ} $\lambda$1549 resonant doublet, respectively. Our ORL C$^{2+}$ abundances are in good agreement with those obtained by \citet{tsamisetal04} but, on the other hand, we obtain C$^{3+}$ and C$^{4+}$ abundances that are somewhat higher than those obtained by \citet{tsamisetal04}. In our case the comparison between UV CELs and optical ORLs is meaningless because  optical and {\it IUE} UV observations do not cover the same volume of NGC\,3918. 

\setcounter{table}{9}
\begin{table}
\begin{small}
\caption{Ionic abundance ratios from permitted lines of C in NGC\,3918$^{\rm a}$.}
\label{C_rls}
\begin{tabular}{cccc}
\noalign{\hrule} \noalign{\vskip3pt}
Mult. & $\lambda_0$ & $I$($\lambda$)/$I$(H$\beta$) ($\times$10$^{-2}$) &   C$^{i+}$/H$^+$($\times$10$^{-5}$) \\
\noalign{\vskip3pt} \noalign{\hrule} \noalign{\vskip3pt}
& & \mc{2}{c}{C$^{2+}$/H$^+$} \\
\noalign{\vskip3pt} \noalign{\hrule} \noalign{\vskip3pt}
6	         & 4267.15  & 0.473$\pm$0.024	& {\bf 45}   \\ 
16.04	& 6151.43	& 0.021$\pm$0.003	& {\bf 47}	  	  \\ 
17.02	& 9903.46$^{\rm b}$	& 0.147$\pm$0.016	& 58	  	  \\ 
17.04	& 6461.95	& 0.049$\pm$0.004	& {\bf 46}	  	  \\ 
17.06	& 5342.38	& 0.025$\pm$0.003	& {\bf 45}  	  \\ 
 & Adopted	 		&			 &{\bf 45$\pm$2} \\
\noalign{\hrule} \noalign{\vskip3pt}
& & \mc{2}{c}{C$^{3+}$/H$^+$} \\
\noalign{\vskip3pt} \noalign{\hrule} \noalign{\vskip3pt}
1	& 4647.42$^{\rm c}$	& 0.235$\pm$0.012	& {\bf 19}         \\ 
16	& 4067.94$^{\rm d}$& 0.113:	& 20  	      \\
	& 4068.91$^{\rm d}$& 0.148:        & 20	       \\
	& 4070.26$^{\rm d}$& 0.193:	& 20 	           \\
	& Sum		& 			& {\bf 20}   \\  
18	& 4186.90	& 0.169$\pm$0.010	& {\bf 21}    \\ 
	& Adopted	& 			&{\bf 20$\pm$2} \\
\noalign{\hrule} \noalign{\vskip3pt}
& & \mc{2}{c}{C$^{4+}$/H$^+$} \\
\noalign{\vskip3pt} \noalign{\hrule} \noalign{\vskip3pt}
8	& 4657.55	& 0.330$\pm$0.023		&  12  	         \\ 
	& 4658.20	&  *		& *  	         \\ 
8.01	& 7726.20	& 0.152$\pm$0.014	&  15 	         \\
	& Adopted	& 			&{\bf13$\pm$2} \\
\noalign{\hrule} \noalign{\vskip3pt}
\end{tabular}
\begin{description}
\item[$^{\rm a}$] Only lines with intensity uncertainties lower than 40\% have been considered (see text).
\item[$^{\rm b}$] Blend with {\fkriii} $\lambda$9902.3 line.
\item[$^{\rm c}$] Blend with unknown line.
\item[$^{\rm d}$] Corrected from the contribution of {\fsii} $\lambda$4068.6 line and {\oii} $\lambda\lambda$4069.62+.89 lines.
\end{description}
\end{small}
\end{table}

\paragraph{N ORLs.}

Several permitted lines of {\nii}, {\niii} and {\niv} are detected in our spectrum. We compute the N$^{2+}$/H$^+$ ratios from {\nii} permitted lines of multiplets 3, 12, 39 and 48, which are mainly excited by recombination 
and are not sensitive to optical depth effects \citep{yliuetal04b}. We assume the recombination coefficients by \citet{fangetal11, fangetal13} except in the case of multiplets 12 and 48, for which \cite{fangetal11,fangetal13} do not compute recombination coefficients and, therefore, we adopt the recombination coefficients by \citet{kisieliusstorey02} for multiplet 12 and \citet{escalantevictor92} for multiplet 48.

Lines belonging to multiplets 1 and 2 are the brighest {\niii} lines detected in our spectrum, but they are not reliable for abundance determinations because they appear
to be excited by the Bowen mechanism \citep{grandi76}. However, other {\niii} lines are detected in our spectrum but the recombination coefficients are only available for the multiplet 18 {\niii} $\lambda$4379.11 line
\citep[see][]{pequignotetal91}. This is a $4f$--$5g$ transition that cannot be excited by mechanisms other than recombination. Unfortunately, this line is probably blended with the {\oiii} $\lambda$4379.6 \AA\ line; as the relative strength of this {\oiii} line with other detected lines belonging to the same multiplet is not available, we can only give an upper limit to the N$^{3+}$/H$^+$ ratio. 

We detect several {\niv} in the spectrum of NGC\,3918, but only a handful of them have available recombination coefficients \citep{pequignotetal91}. These lines give relatively similar results for the N$^{4+}$/H$^+$ ratio and most of them are transitions between high levels and, therefore, are probably excited by pure recombination. The $3s$--$3p$ transition gives very similar results to the other lines and, therefore, recombination is its most probable excitation mechanism. Similarly to the case of C ions, although there are {\it IUE} UV CELs data available for N$^{3+}$ and N$^{4+}$ ions for NGC\,3918, a comparison with our ORL abundances would not be meaningful since optical and UV observations do not cover the same volume of the nebula. 

\setcounter{table}{10}
\begin{table}
\begin{small}
\caption{Ionic abundance ratios from permitted lines of N in NGC\,3918$^{\rm a}$.}
\label{N_rls}
\begin{tabular}{cccc}
\noalign{\hrule} \noalign{\vskip3pt}
Mult. & $\lambda_0$ & $I$($\lambda$)/$I$(H$\beta$) ($\times$10$^{-2}$) & N$^{i+}$/H$^+$  ($\times$10$^{-5}$) \\
\noalign{\vskip3pt} \noalign{\hrule} \noalign{\vskip3pt}
& & \mc{2}{c}{N$^{2+}$/H$^+$}$^{\rm b}$ \\
\noalign{\vskip3pt} \noalign{\hrule} \noalign{\vskip3pt}
3& 5666.64	& 0.019$\pm$0.002       & 10	\\
& 5676.02		& 0.008$\pm$0.002 	    & 8	\\
& 5679.56		& 0.041$\pm$0.003       & 12	\\
& 5710.76           & 0.007$\pm$0.002        & 10          \\
& Sum			       &		    & {\bf 10}	\\ 
12&3994.98        & 0.008$\pm$0.003	   & 8 \\
39&4041.31        & 0.021$\pm$0.003	   & 16 \\
48& 4239.40	& 0.079$\pm$0.006$^{\rm c}$	   & 30  	 \\
& Adopted			& 			&{\bf 10$\pm$1} \\ 
\noalign{\hrule} \noalign{\vskip3pt}
& & \mc{2}{c}{N$^{3+}$/H$^+$} \\
\noalign{\vskip3pt} \noalign{\hrule} \noalign{\vskip3pt}
18& 4379.11$^{\rm d}$	 &  $<$0.160		 &  $<$7  	 \\
& Adopted			& 			&{\bf $<$7} \\ 
\noalign{\hrule} \noalign{\vskip3pt}
& & \mc{2}{c}{N$^{4+}$/H$^+$} \\
\noalign{\vskip3pt} \noalign{\hrule} \noalign{\vskip3pt}
$3s$$^3$S--$3p$$^3$P$_0$&  3483.00& 0.033$\pm$0.013 &  1.6 	 \\
$5g$$^{1,3}$G--$6h$$^3$H$_0$&  4606.33& 0.076$\pm$0.005 & 1.9   	 \\
$6g$$^{1,3}$G--$7h$$^{1,3}$H$_0$&  7581.90& 0.083$\pm$0.019 & 1.2   	 \\
$6h$$^{1,3}$H$_0$--$7i$$^{1,3}$I&  7702.96& 0.025$\pm$0.003 &  1.9  	 \\
& Adopted			& 			&{\bf 1.8$\pm$0.5} \\ 
\noalign{\hrule} \noalign{\vskip3pt}
\end{tabular}
\begin{description}
\item[$^{\rm a}$] Only lines with intensity uncertainties lower than 40\% have been considered (see text).
\item[$^{\rm b}$] For multiplets 3 and 39, recombination coefficients by \citet{fangetal11, fangetal13}. For multiplet 12 recombination coefficients by \citet{kisieliusstorey02}. For multiplet 48, recombination coefficients by \citet{escalantevictor92}.
\item[$^{\rm c}$] The reported flux ratio is the sum of the contribution of three individual lines at $\lambda\lambda$4236.91, 4237.05 and 4241.78 \AA. 
\item[$^{\rm d}$] Blended with {\oiii} $\lambda$4379.6 \AA.
\end{description}
\end{small}
\end{table}

\paragraph{O ORLs.}

\begin{figure}
\begin{center}
\includegraphics[width=\columnwidth]{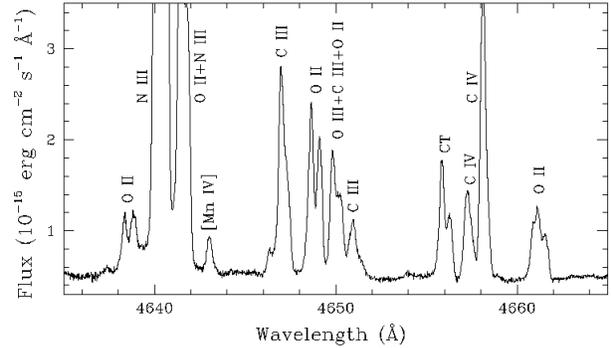}
\caption{Portion of the echelle spectrum of NGC\,3918 showing the zone where the {\oii} multiplet 1 ORLs lie.  Given the high ionization of NGC\,3918, we also detected several permitted lines of {\ciii}, {\civ}, {\niii} and {\oiii} lines, although only some of them are excited by recombination (see text). At $\sim$4656 \AA\ there is a feature produced by charge transfer (CT) in the CCD. }
\label{oii_lines}
\end{center}
\end{figure}

We detect a large number of {\oii} permitted lines in our spectrum. 
\citet{estebanetal04} argue that the best lines to compute O$^{2+}$ abundances from ORLs are those belonging to multiplets 1, 2, 10, 20, and from the $3d$--$4f$ transitions, which are mainly excited by recombination. However, we detect other multiplets that, at the excitation of NGC\,3918, are not expected to be affected by any fluorescence effects.
In Fig.~\ref{oii_lines} we show the region of the multiplet 1 {\oii} ORLs,  which are the brightest {\oii} ORLs in the spectrum of NGC\,3918 (of order 10$^{-3} \times$ $I$({\hb})). The upper levels of the transitions of this multiplet may be affected by departures from local thermodynamic equilibrium (LTE) for densities $n_e$$<$10$^4$ cm$^{-3}$ \citep{tsamisetal03, ruizetal03}. In such a case, the abundances derived from the various lines assuming LTE may differ from each other by large factors. To account for such effect, we apply the non-LTE corrections estimated by \citet{apeimbertetal05}, obtaining abundances from individual lines in good agreement; we also derive the abundance using the sum of all lines of the multiplet following the recipe given by \citet{estebanetal98}. This abundance, which is not affected by non-LTE effects, agrees with those derived from individual lines. 
In Table~\ref{O_rls} we present the values obtained for the individual lines, as well as those derived from the sum of all the lines of a given multiplet. We only consider lines with errors lower than 40\%; for $3d-4f$ transitions, when available, we average lines with errors $<$40\%, otherwise, we average all available lines and quote the final value with an uncertainty higher than 40\%. For 3$s$--3$p$ transitions we use the recombination coefficients assuming LS-coupling by \citet{storey94}. In all the other cases (3$p$--3$d$ and 3$d$--4$f$ transitions) we use the intermediate-coupling 
scheme by \citet{liuetal95}. 

As seen in Table~\ref{O_rls}, the abundances computed from multiplets 1, 5, 10, 20, and $3d$--$4f$ transitions (averaged value) agree within a $\sim$15\% uncertainty. However, multiplets 2 and 19 give abundances that are a factor of $\sim$1.7-2 lower and multiplet 12 give an abundance that is $\sim$10\% higher than the abundances derived from the other lines. \citet{liuetal00} report a similar behavior for multiplet 2 and claim that  departure from case B to case A could resolve this discrepancy. However, considering case A we obtain abundances that are a factor of $\sim$1.4 higher but still too low compared to the other multiplets. Taking into account that multiplet 1 is the most widely used proxy of the O$^{2+}$/H$^+$ ratio from ORLs and that the derived abundance from this multiplet agrees with the one derived from multiplets 5, 10, 20, and $3d-4f$ transitions, we finally adopt the average abundance weighted by the uncertainties of multiplets 1, 5, 10, 20, and $3d-4f$ lines as representative of the O$^{++}$ abundance from ORLs. From this value, we compute an ADF(O$^{2+}$)=1.8$\pm$0.3, which is in good agreement with the value obtained by \citet{tsamisetal04} for NGC\,3918 for their comparison between optical CELs and ORLs (ADF(O$^{2+}$)=1.85). 

We detect several lines of {\oiii} in the spectrum of NGC\,3918. We do not consider the brighest {\oiii} lines (i.~e. those belonging to multiplets 2 and 5) because they are probably excited by other mechanisms than recombination \citep[see][]{grandi76, cleggwalsh85}. We compute the O$^{3+}$/H$^+$ ratio from ORLs from multiplet 8 of {\oiii}, which has contributions from both radiative and dielectronic recombination \citep{liudanziger93a}.

Several {\oiv} lines are detected in our spectrum. We use the $5g$$^2$G--$6h$$^2$H$_0$ {\oiv} $\lambda$4631.89 
and the $6g$$^2$G--$7h$$^2$H$_0$ {\oiv} $\lambda$7677.40 lines to compute the O$^{4+}$/H$^+$ ratio since they are hardly excited by other mechanisms than recombination.

\setcounter{table}{11}
\begin{table}
\begin{small}
\caption{Ionic abundance ratios from permitted lines of O in NGC\,3918$^{\rm a}$.}
\label{O_rls}
\begin{tabular}{cccc}
\noalign{\hrule} \noalign{\vskip3pt}
Mult. & $\lambda_0$ & $I$($\lambda$)/$I$(H$\beta$) ($\times$10$^{-2}$) & O$^{i+}$/H$^+$  ($\times$10$^{-5}$) \\
\noalign{\hrule} \noalign{\vskip3pt}
& & \mc{2}{c}{O$^{2+}$/H$^+$} 	\\
\noalign{\vskip3pt} \noalign{\hrule} \noalign{\vskip3pt}
1$^{\rm b}$& 4638.85		& 0.114$\pm$0.010	& 66	     		\\
& 4641.81			& 0.109$\pm$0.030	& 42	     		\\
& 4649.14			& 0.219$\pm$0.037	& 65	     		\\
& 4650.84			& 0.031:	          	& 17:	     		\\
& 4661.64			& 0.114$\pm$0.027	& 59	     		\\
& 4676.24			& 0.051$\pm$0.024	& 55		      	\\
& Sum				& 			& {\bf 52}	   	\\
2& 4317.14         	& 0.019$\pm$0.003	&  27		      	\\
& 4319.63			& 0.016$\pm$0.003	&  22		      	\\
& 4336.83			& 0.011$\pm$0.003	&  36		      	\\	
& 4345.56    		& 0.022$\pm$0.003    & 28		      	\\
& 4349.43			& 0.053$\pm$0.005    & 26  		      	\\
& Sum				& 			&   27   	   	\\
5& 4416.97         	& 0.029$\pm$0.004	&  {\bf 51}		      	\\
10$^{\rm c}$& 4069.62& 0.135:	         &   53:    		\\
& 4069.89			&	*		& *            		\\	  
& 4072.15                    & 0.133:              &    55:                   \\
& 4075.86			& 0.184:     	&  53:   		\\
& 4078.84                    & 0.021$\pm$0.003   &  57    \\ 
& 4092.93			& 0.018$\pm$0.003	&   51	     	 \\
& Sum				& 			& {\bf 53}	    	 \\ 
12$^{\rm c}$& 3882.19& 0.021$\pm$0.003	&  61	    	\\
19$^{\rm c}$& 4132.80& 0.015$\pm$0.003	& 28	    	\\
& 4153.30			& 0.026$\pm$0.003	&  33     		\\
& Sum				& 			&  31 	\\ 
20$^{\rm c}$& 4110.79& 0.007$\pm$0.002	& 30	    	\\
& 4119.22		& 0.053$\pm$0.005	&  61     		\\
& Sum		& 				& {\bf 57} 	\\ 
$3d$--$4f^{\rm c}$& 4087.15 & 0.015$\pm$0.003     & 51 \\
&4095.64	& 0.007$\pm$0.002		& 32	    		\\	
&4275.55	& 0.036$\pm$0.004		& 60	    		\\	
&4285.69	& 0.011$\pm$0.003		& 50	    		\\	
&4288.82	& 0.005:				& 87:	    		\\	
&4291.25	& 0.010$\pm$0.003		& 55	    		\\	
&4303.61	& 0.004$\pm$0.001		& 60	    		\\
&4303.82	& 0.028$\pm$0.003		& 60	    		\\
&4602.13   &0.008$\pm$0.002		& 44     		\\
&4613.68   &0.003:				& 70:     		\\
& Average			& 			& {\bf 46}		\\ \hline 
& Adopted			& 			&{\bf 52$\pm$7} 		\\ 
\noalign{\hrule} \noalign{\vskip3pt}
& & \mc{2}{c}{O$^{3+}$/H$^+$} 	\\
\noalign{\vskip3pt} \noalign{\hrule} \noalign{\vskip3pt}
8 & 3260.86 	 &  0.286$\pm$0.026	& 18  	 \\
& 3265.33 	 &  0.310$\pm$0.028	& 13   	 \\ 
& 3284.45 	 &  0.024:			& 12   	 \\
& Sum				& 			& {\bf 15} 	\\ 
& Adopted			& 			&{\bf 15$\pm$2} \\ 
\noalign{\hrule} \noalign{\vskip3pt}
& & \mc{2}{c}{O$^{4+}$/H$^+$} 	\\
\noalign{\vskip3pt} \noalign{\hrule} \noalign{\vskip3pt}
$5g$$^2$G--$6h$$^2$H$_0$& 4631.89 	 &  0.153$\pm$0.009	& {\bf 5}   	 \\
20 		& 7677.40 	 &  0.024$\pm$0.003  & {\bf 5}  	 \\
& Adopted			& 			&{\bf 5$\pm$1} \\ 
\noalign{\hrule} \noalign{\vskip3pt}
\end{tabular}
\begin{description}
\item[$^{\rm a}$] Only lines with intensity uncertainties lower than 40\% have been considered (see text).
\item[$^{\rm b}$] Corrected from non-LTE effects (see text).
\item[$^{\rm c}$] Recombination coefficients in intermediate coupling \citep{liuetal95}.
\end{description}
\end{small}
\end{table}

\paragraph{Ne ORLs.}

We detect several permitted lines of {\neii}, belonging to multiplets 1, 7, 39, 55, and 57. To derive Ne$^{++}$ abundances, we adopt the effective recombination coefficients by \citet{kisieliusetal98} for multiplets 1, 7, and 39 and the computations by  Kisielius \& Storey (unpublished), assuming LS-coupling for multiplets 55 and 57. Transitions from multiplet 1 and 55 are probably the result of recombination because they correspond to quartets and their ground level has a doublet configuration \citep{estebanetal04}. Transitions from multiplets 7 and 57 correspond to doublets and give abundances similar (m57) or lower (m7) than those derived from the multiplet 1 and 55 lines; however, given the high uncertainty in the flux of multiplet 7 $\lambda$ 3323.74, we do not include this line in the final average abundance.  Taking into account that the multiplet 55 lines correspond to $3d-4f$ transitions, whose upper levels are unlikely to be populated by fluorescence, it seems that in NGC\,3918 all these lines are excited mainly by recombination.  The multiplet 39 {\neii} $\lambda$3829.77 line corresponds to an intercombination transition (3p$^2$P$^0$--3d$^4$D) and the abundance derived from it agrees with that derived from other multiplets, so this multiplet is probably mainly excited by recombination. Therefore, we adopt the averaged value from all multiplets as representative of the Ne$^{++}$ abundance. From our derived Ne$^{2+}$/H$^+$ ratio from ORLs and CELs, we compute an ADF(Ne$^{2+}$)=2.54$\pm$0.37, which is somewhat higher than that derived for O$^{2+}$, but is within the typical range of PNe. A similar result was found by \citet{tsamisetal04}.

\setcounter{table}{12}
\begin{table}
\begin{small}
\caption{Ionic abundance ratios from permitted lines of Ne in NGC\,3918$^{\rm a}$.}
\label{Ne_rls}
\begin{tabular}{cccc}
\noalign{\hrule} \noalign{\vskip3pt}
Mult. & $\lambda_0$ & $I$($\lambda$)/$I$(H$\beta$) ($\times$10$^{-2}$) & Ne$^{i+}$/H$^+$  ($\times$10$^{-5}$) \\
\noalign{\vskip3pt} \noalign{\hrule} \noalign{\vskip3pt}
& & \mc{2}{c}{Ne$^{2+}$/H$^+$} \\
\noalign{\vskip3pt} \noalign{\hrule} \noalign{\vskip3pt}
1& 3694.22		& 0.060$\pm$0.013	& 16		\\
& 3709.62			& 0.034$\pm$0.013	& 25		\\
& 3766.26			& 0.017$\pm$0.005	& 16		\\
& 3777.14			& 0.015:	                  & 15:		\\
& Sum			&		                  &  {\bf 18}	\\ 
7& 3323.74		& 0.031:			& 8:		\\
39& 3829.77		& 0.013$\pm$0.003	& {\bf 11}		\\
55& 4409.30		& 0.017$\pm$0.003	& {\bf 25}		\\
57& 4428.54		& 0.011$\pm$0.003	& {\bf 25}		\\
& Adopted			& 			&{\bf  16$\pm$2 }               \\ 
\noalign{\hrule} \noalign{\vskip3pt}
\end{tabular}
\begin{description}
\item[$^{\rm a}$] Only lines with intensity uncertainties lower than 40\% have been considered (see text).
\end{description}
\end{small}
\end{table}

\subsection{Total abundances}\label{totalab}

To correct for the unseen ionization stages and then derive the total gaseous abundances of chemical elements in NGC\,3918, it is necessary to adopt a set of ionization correction factors (ICFs). For most elements,  we adopted the ICFs recently developed by \citet{delgadoingladaetal14} from a large grid of photoionization models and applied them to the ionic abundances computed from CELs, but also provided alternative computations when possible. The ICFs derived in \citet{delgadoingladaetal14} were designed for large aperture observations (that cover the whole nebulae) as well as for small aperture observations across the central regions of the nebulae. Our observations avoid the central star but the slit is located sufficiently close to cover different ionization zones in the nebula. Moreover, we obtain similar ionic abundances and ionic fractions [for example, of O$^{++}$/(O$^{+}$+O$^{++}$) and He$^{++}$/(He$^{+}$+He$^{++}$)] than the ones reported in \citet{cleggetal87} and \citet{tsamisetal03b}), whose observations are adequate for these ICFs, and thus, we feel justified the adoption of these ICFs. The total abundances discussed in this section are shown in Table~\ref{total} and the specific criteria adopted in each particular case are discussed in the following. Uncertainties in the total abundances were computed through Monte Carlo simulations that take into account uncertainties in line fluxes, physical conditions and ionic abundances  For C, N, O, Ne, S, Cl, and Ar, we include in the Monte Carlo simulations the uncertainties in the ICFs reported by \citet{delgadoingladaetal14}. 
 
The total C abundance is computed only from ORLs. The ICF proposed by \citet{delgadoingladaetal14} in their Eq. (39) is based on C$^{2+}$ and O$^{2+}$ derived from ORLs and the result obtained is shown in Table~\ref{total}. Alternatively, we also compute the C abundance by adding all the ionic abundances from {\cii}, {\ciii}, and {\civ} ORLs and taking into account the contribution of C$^+$ by re-scaling the C$^+$/C$^{2+}$ obtained by \citet{tsamisetal03b} from CELs. The latter value is only 7\% higher than the value derived from an ICF (see Table~\ref{total}). 

When computing the O abundance from CELs, we apply the ICF by \citet{delgadoingladaetal14} to correct for the unseen high-ionization ions. But we can also compute the O abundance from ORLs. As commented in the previous section, ionic abundances are estimated  for several O ions from ORLs. Therefore, we can estimate the total O abundance from ORLs by adding the observed ionic species, and taking into account the contribution of O$^+$ by scaling the O$^+$/O$^{2+}$ obtained from CELs. Alternatively, we can also compute the O total abundance from O$^{2+}$ obtained from ORLs and the scaled O$^+$ abundance from CELs and the ICF provided by \citet{delgadoingladaetal14}. As shown in Table~\ref{total}, there is an excellent agreement between both values, indicating that the ICF seems to estimate correctly the contribution of high ionization ions to the total O abundance.

Thanks to the high ionization of NGC\,3918, we detect lines of {\fneiii}, {\fneiv}, and {\fnev} with a high S/N ratio (see Fig.~\ref{profiles}). We need to correct for the contribution of Ne$^+$ to the total abundance. Therefore, we compute the total abundance in two ways: i) by adding the ionic chemical abundances, and ii) by using a classical ICF scheme that corrects the observed ionic abundances for the unseen ionization species. We use the ICFs given by Eqs. (17) and (20) of \citet{delgadoingladaetal14}, which give the total Ne abundance when only {\fneiii} lines are observed and both {\fneiii} and {\fnev} lines are observed. All the results are shown in Table~\ref{total}. The abundance obtained from the sum of the ionic species agree within the uncertainties with that derived using the ICF when {\fneiii} and {\fnev} lines are detected, but it is $\sim$0.1 dex higher than that derived using only {\fneiii} lines. Because auroral {\fneiv} lines are quite sensitive to the electron temperature, these ICFs do not use them. Considering all these factors, we adopt as a representative Ne abundance the value given by Eq. (20) of \citet{delgadoingladaetal14}, which is 8.01$\pm$0.06.

For Cl, we detect {\fclii}, {\fcliii}, and {\fcliv} lines and, therefore, we use the ICF recommended by \citet{delgadoingladaetal14} when all these lines are observed (their Eq. 32), which takes into account the contribution of high Cl ionization states.

Similarly to Ne, we detect several ionization stages of Ar in the spectrum of NGC\,3918 (see Fig.~\ref{profiles}). We compute the total Ar abundance by adding the ionic chemical abundances and by using the ICF proposed by \citet{delgadoingladaetal14} when only {\fariii} lines are detected (Eq. 36). The uncertainties associated with the ICF reported by these authors are very high (see their Table 3 and Figure 13). Note that these uncertainties are likely overestimated since they are derived from the maximum dispersion in the ICF given by the grid of photoionization models. 
This is the reason why we obtain high errors on the Ar/H determination based on this ICF (see Table~\ref{total}). Given that the ICFs for Kr and Xe are based on Ar abundances, we have tried to find a value for Ar/H with lower uncertainties. We select from the 3MdB models \citep{morissetetal15} the ones corresponding to the PNe\textunderscore2014 reference \citep[used by ][ to compute their ICFs]{delgadoingladaetal14}, adding as a selection criteria the fact that the models must reproduce simultaneously the Ar$^{2+}$/Ar, Ar$^{3+}$/Ar, Ar$^{3+}$/Ar, and O$^{2+}$/O ionic fractions (within $\pm$5\%). These models then reproduce well the ionization structure of the nebula. The total Ar/H abundance for these models is found to be very close to the sum Ar$^{2+}$/H + Ar$^{3+}$/H + Ar$^{3+}$/H, with an error being less than 2\%. Therefore, we will use from now on the sum of these three ionic abundances as representative of the total Ar abundance.

\begin{figure*}
\begin{center}
\includegraphics[width=\textwidth]{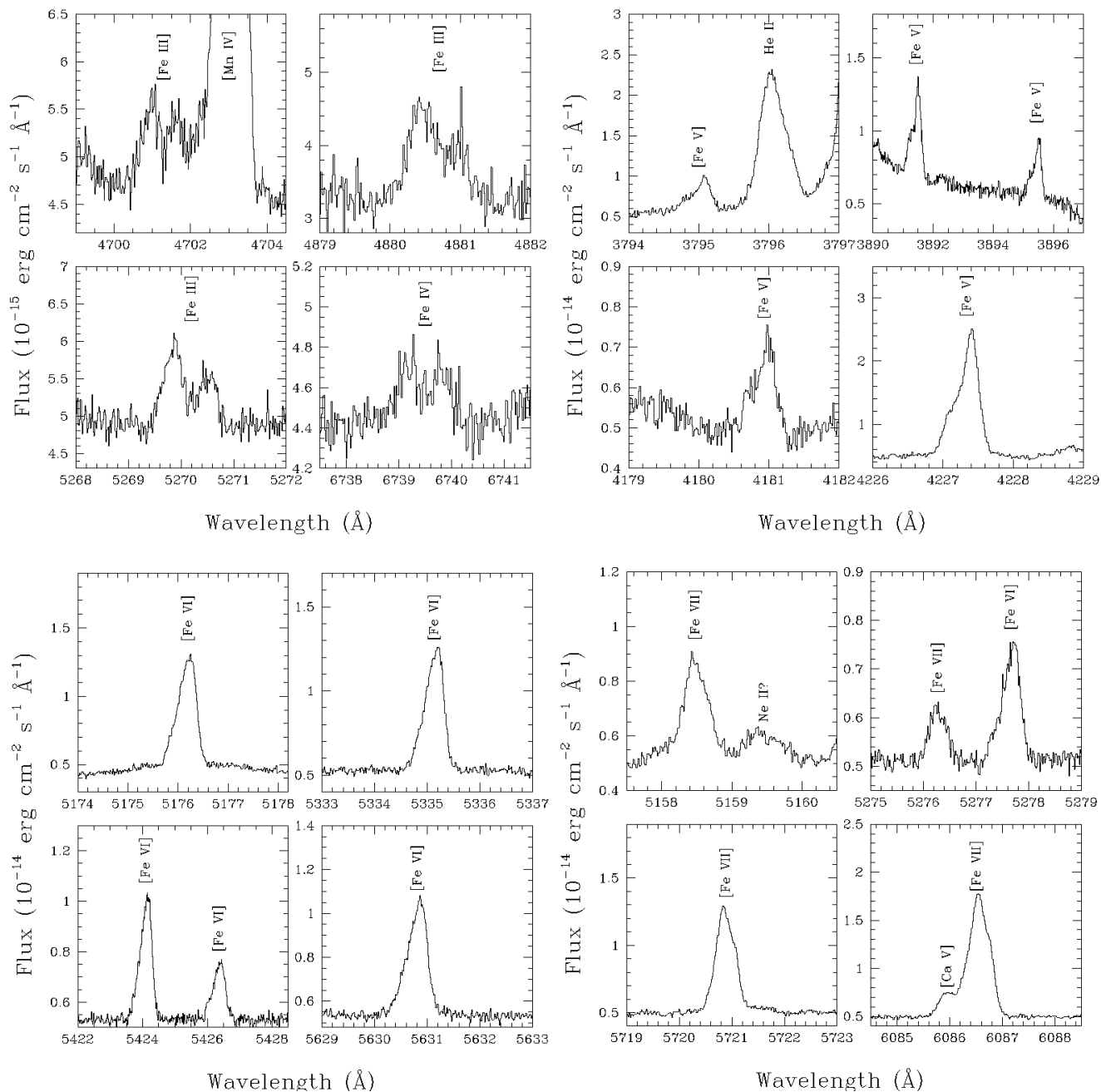}
\caption{Portions of the echelle spectrum of NGC\,3918 showing lines of different ionic species of Fe.}
\label{iron_lines}
\end{center}
\end{figure*}

As can be seen in Table~\ref{ionic}, we measure several lines of multiple ionization stages of Fe. In Fig.~\ref{iron_lines} we show some of the lines used to compute ionic chemical abundances. The {\ffeiii} lines are usually the brightest iron lines in photoionized nebulae. However, as is shown in Fig~\ref{iron_lines}, this is not the case for NGC\,3918, which is an extremely high ionization PN and presents several bright {\ffev}, {\ffevi} and {\ffevii} lines. Taking this into account, we use two different approaches to compute the total Fe abundance: i) in the first one, we use the correction scheme suggested by \citet{rodriguezrubin05}, which is based on the detection of {\ffeiii} lines and two ICFs, one from photoionization models (their Eq. 2) and one from an observational fit that takes into account all the uncertainties in the atomic data involved in the calculations and that allows us to constrain real iron abundances (Eq. 3); ii) in the second approach, we take advantage of the high ionization degree of the PN and simply compute the total abundance as the sum of all the observed ionic species. The results are shown in Table~\ref{total}. As can be seen, the sum value is in very good agreement with the value derived from Eq. (2) of \citet{rodriguezrubin05}. \citet{delgadoingladarodriguez14} also find several objects with high-ionization Fe ions. In a similar exercise to ours, they compare the sum of the ionic Fe abundances with the results obtained from the ICFs, but their results do not allow them to conclude which ICF works better. It must be taken into account that their data are more limited than our data both in spectral resolution, which is lower and, thus, prevents deblending some of the lines from other close-by features, and in the S/N ratio of the high-ionization Fe lines, which is also lower. We agree with their conclusion that the inherent limitations of ionization correction schemes are responsible for the different behaviours found and that high-resolution, high-S/N ratio data are needed to improve the ionization correction scheme or to define a new and better one, especially for high ionization PNe. 

To estimate the correction for the unseen ions of Ca, K and Na, for which no published ICFs exist, we compute a simple spherical photoionization model of the PN with Cloudy \citep[v. 13.03;][]{ferlandetal13}, ionized by a blackbody with $T_{eff}$=190 kK.  
The inner and outer radius are set to log($r_{in}$)=15.0 and log($r_{out}$)=17.16 (in cm);
at an assumed distance of 1 kpc, the projected radius equals the observed optical radius (19$''$). Using pyCloudy \citep[][]{morisset13} Cloudy's output has been corrected to account for the slit used to observe the nebula (9.35$''$ $\times$ 1$''$: see Section~\ref{obsred}).
The density, chemical composition and total {\hb} flux are set by design. 
The model thus computed reproduces the observed line ratios
and the general ionization structure as given by {\oi} through {\oiv} and {\neiii} through {\nev}. 
According to the model, Na$^{3+}$ accounts for 35\% of the total sodium; K$^{3+}$ and K$^{4+}$ together account for the 46\% of the total potassium; and Ca$^{4+}$ accounts for 17\% of the total calcium. This leads to an ICF of $\sim$2.9 for Na, $\sim$2.2 for K and $\sim$5.8 for Ca.
The validity of the model is supported by the nice agreement between the  K$^{3+}$/K$^{4+}$ ratios predicted by our model (1.04) and the one derived from the observations (1.10$\pm$0.25).

An alternative method to compute the abundances of these three elements is the ICF scheme by \citet{delgadoingladaetal14}, based on the public database of photoionization models 3MdB developed by \citet{morissetetal15}. 
We preliminarly verify the overall agreement of our model with the predictions by \citet{delgadoingladaetal14} by comparing the ICFs of that paper with those returned by the model; the differences between both methods are 0 dex, -0.04 dex, -0.05 dex and +0.04 dex for He/H, Ne/H, Ne/O, and C/O and indicate a substantial agreement between the two methods.   
Based on the values of O$^{2+}$/(O$^{+}$+ O$^{2+}$) and He$^{2+}$/(He$^{+}$+He$^{2+}$) of the nebula, a rough estimate gives us the following ICFs: $\sim$4.0 for Na, $\sim$3.2 for K and $\sim$10.0 for Ca. 

Given the overall agreement between the two methods, we adopt as a representative value of the total abundance for each atomic species the average of the values returned by both methods. The results are shown in Table~\ref{total}.

The gaseous abundances of Ca, K, and Na relative to hydrogen in the interstellar medium are significantly lower than solar \citep{morton74}, which is generally attributed to depletion into dust grains. The reported abundances of K and Na in PNe are about 1 dex lower than solar and the Ca abundances are between $\sim$0.8 and $\sim$2.5 dex lower than solar \citep{aller78, alleretal81, allerczyzak83, bohigasetal13, shieldsetal81, keyesetal90}. The abundances of Ca, K, and Na obtained here are 1.6, 0.6, and 0.3 dex lower than solar respectively, suggesting a significant depletion of these elements (especially of Ca) into dust grains in NGC\,3918.   

The high depletion factors in NGC\,3198 could be related to the dominant chemistry in the PN. Both the value of C/O ($\gtrsim$1, this work) and the broad feature at 30 $\mu$m found in the nebula \citep{delgadoingladarodriguez14} and often associated with MgS, reveal a carbon-rich environment. In the particular case of iron, \citet{delgadoingladarodriguez14}  found that the highest depletion factors are found in carbon-rich PNe.

\citet{sterlingetal07} constructed the first ionization correction scheme for Kr, using approximate atomic data. These authors propose ICFs based on detailed photoionization models and on the detection of multiple Kr ions in the optical and K-band. Very recently, \citet{sterlingetal15} incorporated new photoionization and recombination data into the modeling code Cloudy \citep{ferlandetal13} . They compute a large grid of photoionization models to determine a new set of ICFs for calculating the Kr elemental abundance. This new set provides different ICFs depending on the detection of different Kr ionic lines: {\fkriii}, {\fkriv}, and {\fkrv} and therefore, is perfectly fitted for our data. We note that the results obtained by using the different ICFs proposed by \citet{sterlingetal15} display a relatively low dispersion, suggesting that the formulae proposed are quite robust. In Table~\ref{total} we show the total Kr abundances obtained by using the average of the ICFs proposed by \citet{sterlingetal07} and the average of the ICFs obtained with Eqs. (4), (5) and (6) (which are the ones that use more than one ionization stage) of \citet{sterlingetal15}. We rely more on the second set of ICFs because they were computed from models that incorporated \textit{ab initio}
atomic data calculations\citep{sterling11, sterlingstancil11}. \citet{sharpeeetal07} also propose an ICF based on the similarity between the ionization potentials of the noble gases. This ICF scheme gives a result which is very similar (0.01 dex higher) to the sum of the observed ionic species, but is much lower than the other ICF schemes. 

There are no reliable ICFs for Xe and Rb in the literature. \citet{sharpeeetal07} propose a correction for Xe based on the similarity in the ionization potentials of noble gases. We used a similar scheme assuming that Xe/Ar$\approx$(Xe$^{2+}$ + Xe$^{3+}$)/(Ar$^{2+}$ + Ar$^{3+}$). This scheme yields an abundance that is slightly lower than the sum of the ionic abundances. Therefore, for Xe and Rb, the total abundance is simply computed as the sum of all the observed ionic species. This gives us a lower limit to the total abundance. 

\setcounter{table}{13}
\begin{table*}
\caption{Total abundances.}
\label{total}
\begin{tabular}{lcccccccc}
\noalign{\smallskip} \noalign{\smallskip} \noalign{\hrule} \noalign{\smallskip}
Ion 	&  \multicolumn{4}{c}{12 + log(X/H)} & \multicolumn{2}{c}{C87$^{\rm a}$}		& \multicolumn{2}{c}{T03, T04$^{\rm b}$}		\\
         & CELs sum & CELs ICF  & ORLs sum & ORLs ICF & CELs & ORLs & CELs & ORLs\\
\noalign{\smallskip} \noalign{\hrule} \noalign{\smallskip}
He		& --- & --- &	11.01$\pm$0.01		& ---  & --- & 11.03 &--- & 11.00	\\[3pt]
C 		& --- & --- &    8.93$\pm$0.04$^{\rm c}$ &   8.90$\pm$0.09$^{\rm d}$ & --- & 8.90 & 8.64 & 8.88	\\[3pt]
N    		& --- &  8.12$\pm$0.09 & --- & --- & 8.18 & --- & 8.02 & ---	\\[3pt]
O    		& --- &  8.67$\pm$0.06 & 8.90$\pm$0.05$^{\rm e}$ & 8.90$\pm$0.11$^{\rm f}$ & 8.70 & --- & 8.86 & 9.09	\\[3pt]
Ne              & 8.04$\pm$0.06 & 7.95$\pm$0.06/8.01$\pm$0.06$^{\rm g}$ & --- & 8.44$\pm$0.08 & 8.08 & --- & 7.97 & ---	  \\[3pt]
Na   		& --- &  6.07: & --- & --- & --- &  --- & --- & ---	\\[3pt]
S    		&  --- & 6.81$^{+0.06}_{-0.11}$ & --- & --- & 7.20 & --- & 6.70 & --- \\[3pt]
Cl   		& --- &  5.17$\pm$0.10 & --- & --- & --- &  --- & 5.11 & ---	\\[3pt]
Ar		& 6.43$\pm$0.03 & 6.40$^{+0.08}_{-0.16}$$^{\rm h}$ & --- & --- & 6.30 & --- & 6.24 & --- \\[3pt]
K		& --- &  4.85: & --- & --- & --- &  --- & --- & ---	\\[3pt]
Ca		& --- &  4.66: & --- & --- & --- &  --- & --- & ---	\\[3pt]
Fe		&  4.95$\pm$0.07 & 4.84$\pm$0.08/4.48$\pm$0.10$^{\rm i}$ & --- & --- & 5.57 & --- & --- & --- \\[3pt]
Se		& --- & 3.52$^{+0.15}_{-0.23}$ & --- & --- & --- & --- & --- & --- \\[3pt]
Kr		& 3.80$\pm$0.04 & 3.97$\pm$0.12/3.89$\pm$0.06$^{\rm j}$ & --- & --- & --- & --- & --- & --- \\[3pt]
Rb              &  2.51$^{+0.14}_{-0.20}$ & --- & --- & ---  & --- & --- & --- & --- \\[3pt]
Xe              &  2.49$\pm$0.07  & 2.48$\pm$0.08 & --- & ---  & --- & --- & --- & --- \\[3pt]
\noalign{\smallskip} \noalign{\hrule} \noalign{\smallskip}
\end{tabular}
\begin{description}
\item[$^{\rm a}$] \citet{cleggetal87}.
\item[$^{\rm b}$] \citet{tsamisetal03b, tsamisetal04}.
\item[$^{\rm c}$] Sum of C ionic species from ORLs, with C$^+$ estimated from the CEL C$^+$/C$^{2+}$ ionic ratio by \citet{tsamisetal03b} (see text).
\item[$^{\rm d}$] From {\cii} ORLs and the ICF by \citet{delgadoingladaetal14}.
\item[$^{\rm e}$] Sum of O ionic species from ORLs, with O$^+$ estimated from the CEL O$^+$/O$^{2+}$ ionic ratio (see text).
\item[$^{\rm f}$] From {\oii} ORLs, O$^+$/O$^{2+}$ from CELs, and the ICF by \citet{delgadoingladaetal14}.
\item[$^{\rm g}$] ICF from Eq. (17)/Eq. (20) of \citet{delgadoingladaetal14}.
\item[$^{\rm h}$] ICF from Eq. (36) of \citet{delgadoingladaetal14}.
\item[$^{\rm i}$] Sum of all ionic abundances/ICF from Eq. (2)/Eq. (3) by \citet{rodriguezrubin05}.
\item[$^{\rm j}$] Average from the ICFs given by \citet{sterlingetal07}/Average from the ICFs given by Eqs. (4), (5) and (6) of \citet{sterlingetal15}.
\end{description}
\end{table*}

\subsection{Comparison with elemental abundances in the literature}

The first attempt to compute chemical abundances in NGC\,3918 was made by \citet{torrespeimbertpeimbert77} from optical photoelectric spectrophotometry using a wide long-slit. \citet{torrespeimbertetal81} obtain $IUE$ low dispersion spectra and construct ionization structure models to derive physical conditions and chemical content of the PN. \citet{penatorrespeimbert85} derive the physical conditions and chemical abundances from high-dispersion $IUE$ spectra. The most detailed chemical analysis of NGC\,3918 is the one by \citet{cleggetal87}, who use a combination of high- and low-resolution $IUE$ and optical long-slit spectra covering different positions in the nebula. Using UV, optical and far-infrared spectrophotometric data covering the whole nebula, \citet{tsamisetal03b} and \citet{tsamisetal04} study the chemical content of NGC\,3918 derived from CELs and ORLs. Finally, \citet{ercolanoetal03b} construct a realistic photoionization model of NGC\,3918 using a three-dimensional Monte Carlo photoionization code.

From Table~\ref{total} it is clear that the agreement between our total abundances with those elements in common with \citet{cleggetal87} is very good. For He, C, N, O, and Ne, their derived total abundances agree with ours within 0.05 dex. For Ar, the difference is slightly larger (between 0.1 and 0.2 dex depending on the ICF assumed). The abundances derived by these authors are computed through a tailored photoionization model to constrain UV and optical observations. In \citet{cleggetal87} computations, only the S/H and Fe/H ratios are significantly higher, by 0.4 dex and more than 0.6 dex, respectively.  In Table~\ref{total} we also include the abundances obtained by \citet{tsamisetal03b} from UV, optical and far-infrared CELs of C, N, O, S, Cl, Ne and Ar and those derived by \citet{tsamisetal04} from C and O ORLs. In this case the agreement between C ORLs abundances is also very good. In general, the abundances agree within 0.1-0.15 dex, with a larger scatter for O ($\sim$0.2 dex). In the case of CELs O abundances, the difference can be attributed mainly to the fact that they compute a high O$^{3+}$/H$^+$ ratio from the UV O~{\sc iv}] $\lambda$1401 line, which is very sensitive to both temperature and extinction effects and, to a lesser extent, to the fact that they use the classical ICF scheme of \citet{kingsburghbarlow94}, which is based on a small grid of photoionization models. In the case of RL abundances, the differences are due to the fact that they scaled O$^{3+}$/O$^{2+}$ from CELs, which is quite uncertain and is a factor of $\sim$4 higher than the ratio obtained from ORLs.

\section{Discussion}\label{discuss}

\subsection{{\emph s}-process enrichments of {\emph n}-capture elements}

The total abundances of Se, Kr, Rb, and Xe can be used to establish if the central star has undergone a substantial \emph{s}-process nucleosynthesis and convective dredge-up. As discussed in \citet{sterlingdinerstein08}, the enrichment must be measured against a reference element which is neither enriched nor depleted in the object studied; for a metal-rich, non-type-I PN such as NGC\,3918, a suitable element seem to be O; in case of a type-I PNe, O can be affected by nucleosynthesis and then we use Ar as a reference \citep[see][]{sterlingdinerstein08}. 
In order to establish which level of [X/O], where X is a {\emph n}-capture element, counts as significant, it is necessary to take into account the primordial scatter in the abundances of the progenitor gas. Based on stellar data by \citet{travaglioetal04} and \citet{burrisetal00}, \citet{sterlingdinerstein08}
estimate that \emph{s}-process enrichment is statistically likely to have taken place if the [X/O] ratio is found to be in excess of 0.2-0.3 dex. 

It should be noted that \citet{delgadoingladaetal15} have recently found evidence of O enrichment in a group of Galactic PNe with carbon-rich dust, including NGC\,3918. This result suggests that the real {\emph s}-process element enrichment in this PN is likely higher. However, in order to compare our results with those found by \citet{sterlingetal07} and \citet{sterlingetal15} we still consider O as the metallicity proxy for these PNe.

The [Se/O], [Kr/O], [Xe/O] and [Rb/O]\footnote{ [X/O]=log(X/O)-log(X/O)$_{\odot}$. Solar abundance references are taken from \citet{asplundetal05b}} ratios determined for NGC\,3918 from our data are shown in Table~\ref{sproc_enrich}. If we
adopted the more recent solar composition of \citet{asplundetal09}, the derived abundances
would change by less than 0.06 dex, with the exception of [Rb/O] which would be 0.11 dex
larger. In the calculation, we assume the solar abundances by \citet{asplundetal05b} for the sake of a direct comparison with the results by \citet{sterlingdinerstein08}. For Kr we obtain [Kr/O]=0.57$\pm$0.18, a value well above the threshold assumed by \citet{sterlingdinerstein08}. We can therefore conclude that Kr is strongly enriched in NGC\,3918.
It is interesting to note that our Kr data, which include all the relevant ionization stages, make it possible to test the predictions of theoretical ICFs; the results support the ICFs proposed by \citet{sterlingetal15}. 

For Se, owing to the difficulties detecting and measuring the line, the value we obtain, i.~e. [Se/O] = 0.18$^{+0.17}_{-0.22}$ is very uncertain. Taking this result at face value, we find that Kr is more enriched than Se in NGC\,3918 ([Kr/Se]=0.42$^{+0.21}_{-0.26}$), in agreement with what is observed by \citet{sterlingdinerstein08} in a large sample of PNe with both Kr and Se emission ([Kr/Se]=0.5$\pm$0.2) and with the predictions of theoretical models of {\emph s}-process nucleosynthesis \citep{gallinoetal98, gorielymowlavi00, bussoetal01, karakasetal09}.

The values obtained for Rb and Xe, in turn, are only marginally positive ([Rb/O]$>$0.17$^{+0.17}_{-0.22}$; [Xe/O]$>$0.21$\pm$0.11), so they do not allow us to draw any definite conclusion.
It must be stressed, though, that these values are only lower limits to the real
abundances, so we cannot exclude that these elements are actually
enriched. 

These two elements lie on opposite sides of the bottleneck in the \textit{s}-process path corresponding to
isotopes with the magic neutron number $N=50$. This bottleneck occurs because species
with closed neutron shells have very small {\emph n}-capture cross sections, leading to peaks in the
element-by-element distribution of \textit{s}-process enrichments \citep[e.~g.][]{kappeleretal11}. In stellar spectroscopy, \textit{n}-capture elements produced by the \textit{s}-process
are often categorized as `light-\textit{s}'' or ``heavy-\textit{s}'' elements if one of their isotopes
has a magic number of $N=50$ (e.g., the \textit{s}-process peak near Zr) or $N=82$ (the
peak near Ba), respectively \citep[e.~g.,][]{luckbond91}. The abundance ratio [hs/ls] is a
measure of the neutron exposure (the time-averaged neutron flux experienced by Fe-peak
nuclei) in AGB stars. Both Kr and Xe have isotopes with a magic number of neutrons, and
can be used as nebular proxies for light-\textit{s} and heavy-\textit{s} elements, respectively.
The ratio [Xe/Kr]$\geq -0.39$ corresponds to a neutron exposure typical of modestly
subsolar metallicity AGB stars, according to the models of \citet[][see their Figures 3 and
4]{bussoetal01}. While this abundance ratio is a lower limit, it appears to be somewhat lower than the average $[$Xe/Kr$]\approx 0.0$ value found by \citet{sharpeeetal07} in their sample of PNe. 
Theoretical studies have shown that a spread in the sizes of $^{13}$C pockets and in average neutron exposures are needed, even in stars with the same mass and metallicity, in order to reproduce observations of s-process enrichments in AGB stars \citep[e.~g.,][]{bussoetal01, herwig05, karakaslattanzio14}.  \citet{herwigetal03} showed that the interplay between convective overshoot and rotationally-induced mixing can naturally lead to a spread in neutron exposures in AGB stars of a given mass and metallicity. 
In order to draw more robust conclusions from these results, advances on two fronts are needed.  First, a reliable ICF scheme is needed for Xe.  This requires photoionization and recombination data such as \citet{sterlingwitthoeft11} and \citet{sterling11} produced for Se and Kr.  Secondly, AGB nucleosynthesis model studies are needed to interpret [Xe/Kr] ratios and their relation to the [hs/ls] index used in studies of s-process nucleosynthesis in AGB stars.

Though only a lower limit, the modestly enhanced [Rb/O] value suggests that the reaction $^{13}$C($\alpha$, n)$^{16}$O was the neutron source in the AGB progenitor of NGC\,3918, and not $^{22}$Ne($\alpha$, n)$^{25}$Mg which operates more efficiently in intermediate-mass AGB stars \citep{vanraaietal12, karakasetal12}. In particular, the high neutron density of the $^{22}$Ne source opens branchings in the {\emph s}-process path that lead to large enrichments of Rb relative to other {\emph n}-capture elements, unlike the $^{13}$C neutron source, which produces lower neutron densities. Therefore the modest Rb enrichment relative to Kr and Xe indicates that the progenitor mass of NGC\,3918 is between 1.5 M$_{\odot}$ \citep[the approximate lower limit for convective dredge-up to occur;][]{stranieroetal06} and 5--6 M$_{\odot}$ \citep{karakasetal12}. The upper limit to the progenitor mass is consistent with the abundances of other species in NGC\,3918, in particular N/O and C/O. NGC\,3918 is not a Peimbert Type I PN \citep{peimbert78}, and thus its progenitor mass is expected to be less than 3--4 M$_{\odot}$. The analysis of {\emph n}-capture element abundances confirms this expectation, and additionally places a lower limit to the progenitor mass of this PN.

A more definite statement regarding Rb and Xe would require either observing all the relevant ionization stages or correcting the observed ones with an ICF.
Although corrections for unseen ions of Se and Kr exist \citep{sterlingetal07, sterlingetal15}, no ICF expressions are available for Xe and Rb as yet, due to a lack of the atomic data required to derive such corrections. However,
theoretical and experimental efforts to produce these atomic data are ongoing \citep{kerlinetal15}, after which photoionization modeling can be used to determine analytical ICF
formulae \citep[e.~g.,][]{sterlingetal15}. As already noted above, photoionization models are especially productive if used in combination with data such as those presented in this paper, since the detection of several ionization stages for each element makes it possible to test the predictions of theoretical ICFs. 

\setcounter{table}{14}
\begin{table}
\caption{Total abundance ratios} 
\label{sproc_enrich} 
\centering
\begin{tabular}{lc} 
\hline\hline 
Ratio 	& [X/O]$^{\rm a}$ 	\\
\hline 
[Se/O]		  &  0.18$^{+0.17}_{-0.22}$ 	 \\[3pt]
[Kr/O]$^{\rm b}$  &  0.60$\pm$0.13	\\ [3pt]
[Rb/O]$^{\rm c}$  &  $>$0.17$^{+0.17}_{-0.22}$	\\ [3pt]
[Xe/O]$^{\rm c}$  &  $>$0.21$\pm$0.11	\\ [3pt]
\hline
\end{tabular}
\begin{description}
\item[$^{\rm a}$] [X/O]=log(X/O)-log(X/O)$_{\odot}$. Solar abundances from \citet{asplundetal05b}.
\item[$^{\rm b}$] ICF for Kr is an average of the ICFs proposed by \citet{sterlingetal15} from Eqs. (4), (5) and (6).
\item[$^{\rm c}$] Lower limits. Total Rb and Xe from the sum of the ionic abundances.
\end{description}
\end{table}

\subsection{Correlation of {\emph s}-process enrichments with the C/O ratio}

\begin{figure}[h]
\center
\includegraphics[width=9cm]{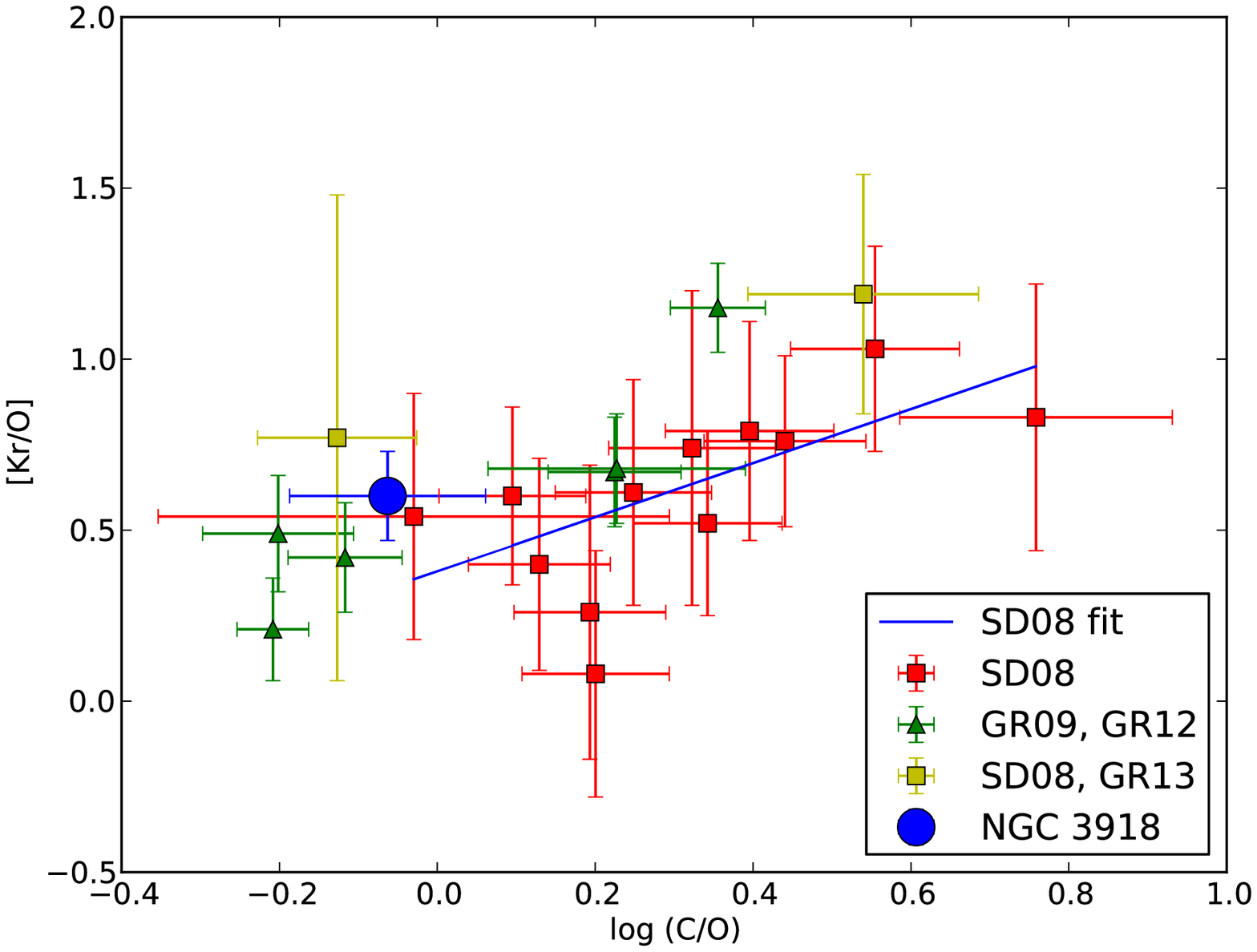}
\includegraphics[width=9cm]{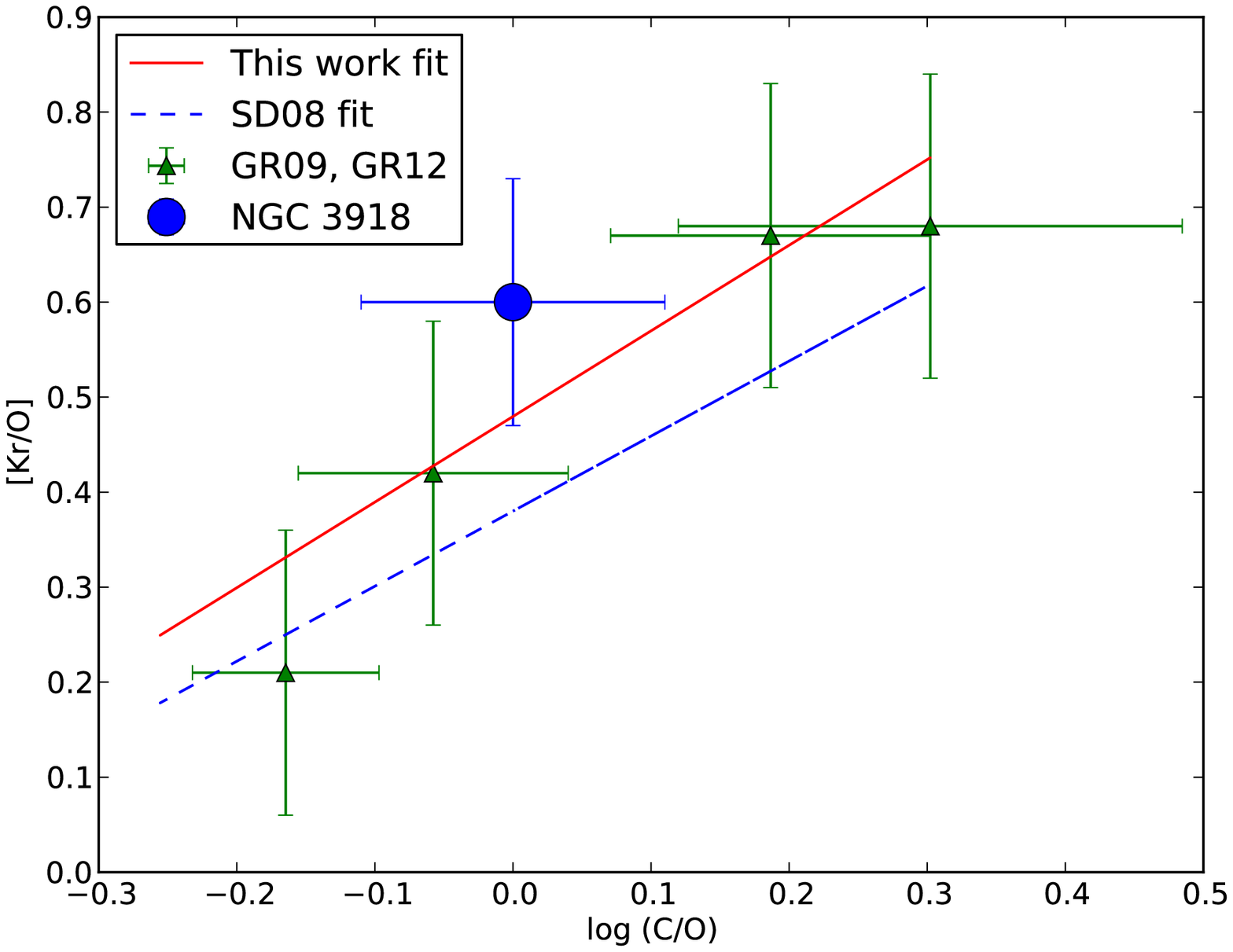}
\caption{[Kr/O] $vs.$ log(C/O) correlation. Upper panel: red squares represent the data by \citet{sterlingdinerstein08} with C/O ratios computed from UV  forbidden lines. Green triangles are data for objects of the sample by \citet{garciarojasetal09, garciarojasetal12, garciarojasetal13} with more than one ionization stage of Kr detected (see text); yellow squares are two objects with [Kr/O] ratios from \citet{sterlingdinerstein08} and log(C/O) determinations from ORLs \citet{garciarojasetal13}. The blue dot represent the result obtained for NGC\,3918. For NGC\,3918 and the objects of the sample of \citet{garciarojasetal09} and \citet{garciarojasetal12} where the C/O ratios are computed from optical recombination lines and using the ICF scheme by \citet{kingsburghbarlow94}. The least squares fit to the \citet{sterlingdinerstein08} data is shown as a blue line. Lower panel: NGC\,3918 and objects from the sample of \citet{garciarojasetal12} with measured {\fkriii} and {\fkriv} lines. In this case, we apply the ICF scheme by \citet{delgadoingladaetal14} to compute the total C and O abundances. A least squares fit to the data is shown as a red line. For comparison we also show the fit to the \citet{sterlingdinerstein08} data.}
\label{KrO_CO}
\end{figure}

\begin{figure}[h!]
\begin{center}
\includegraphics[width=\columnwidth]{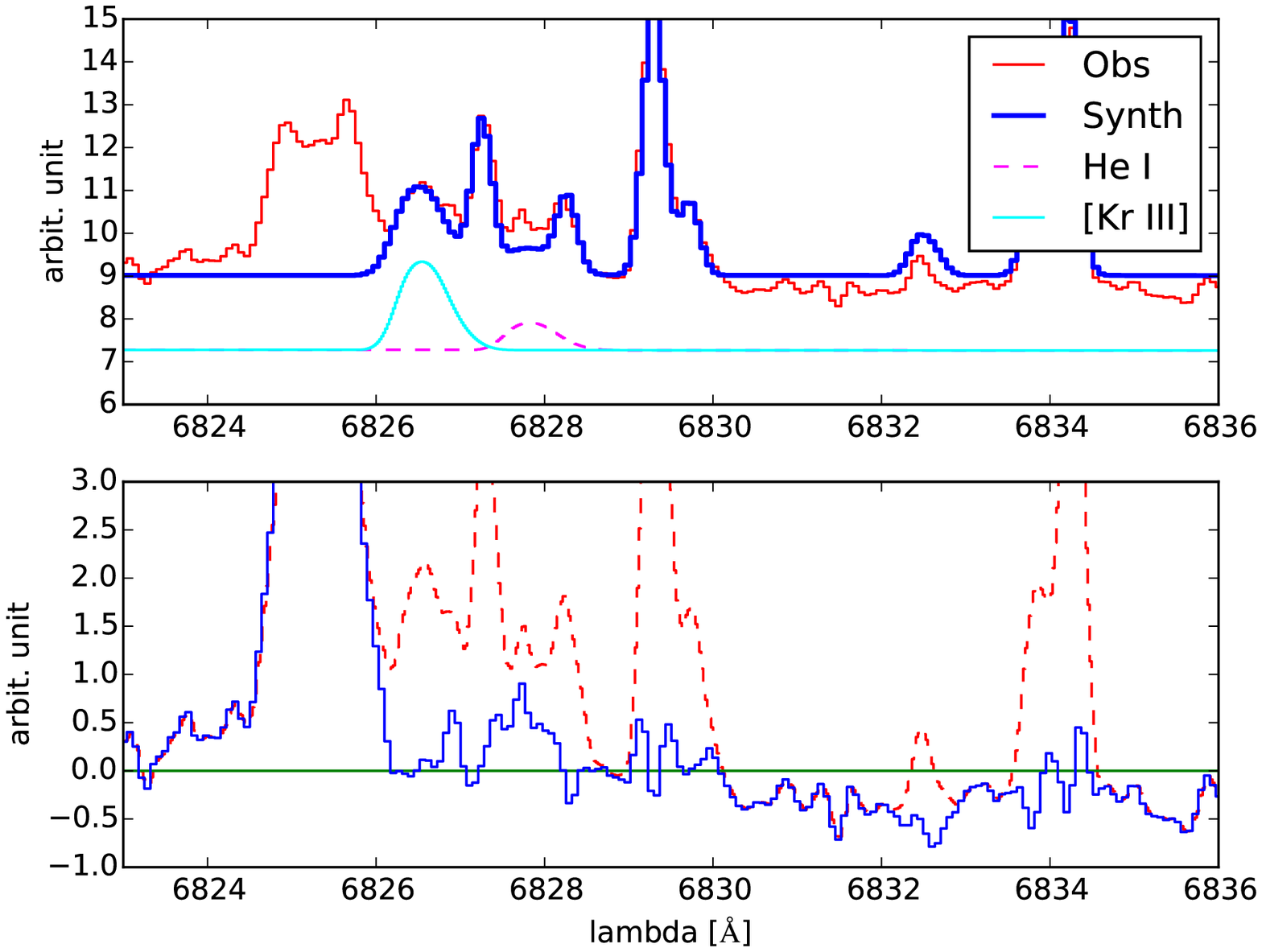}
\includegraphics[width=\columnwidth]{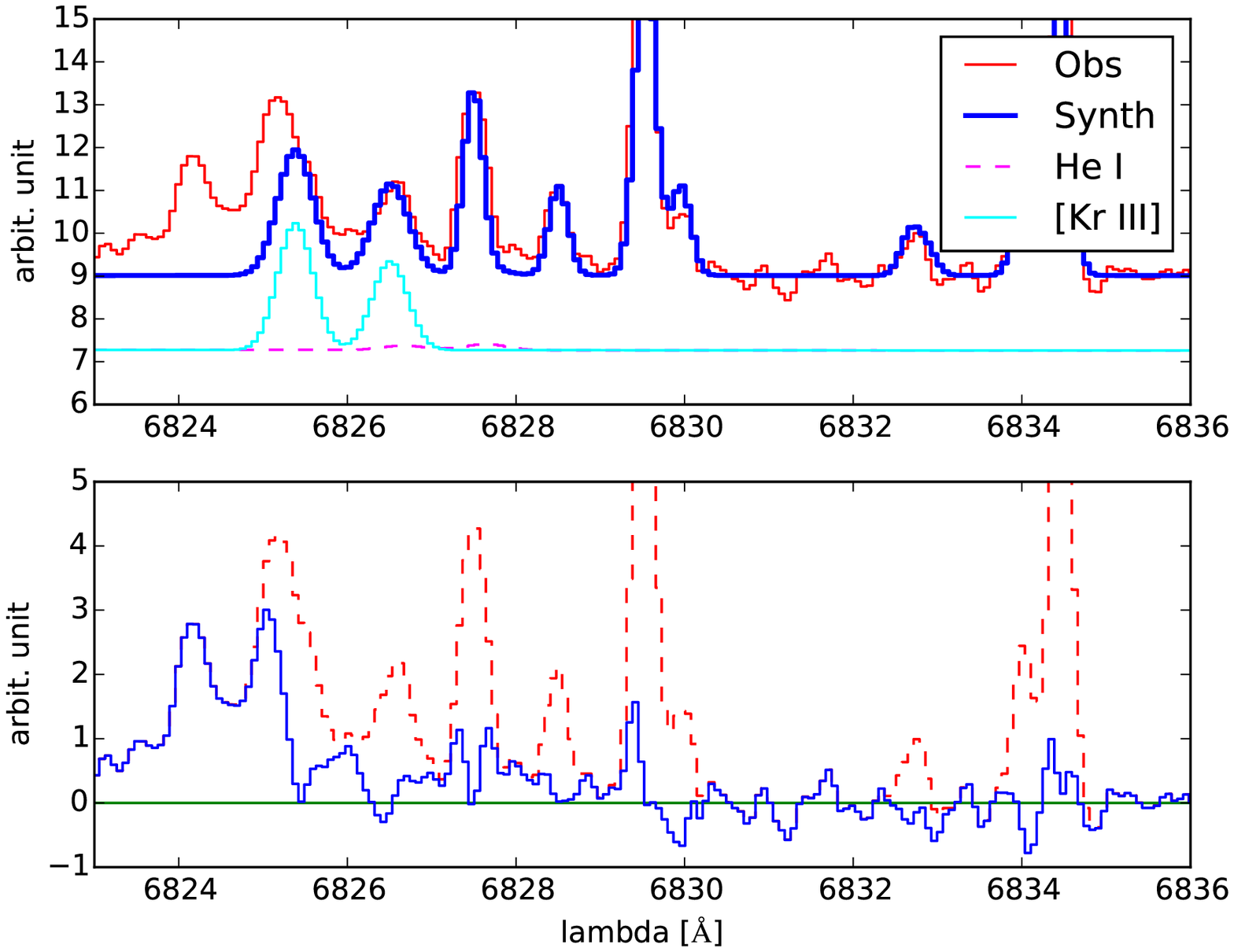}
\includegraphics[width=\columnwidth]{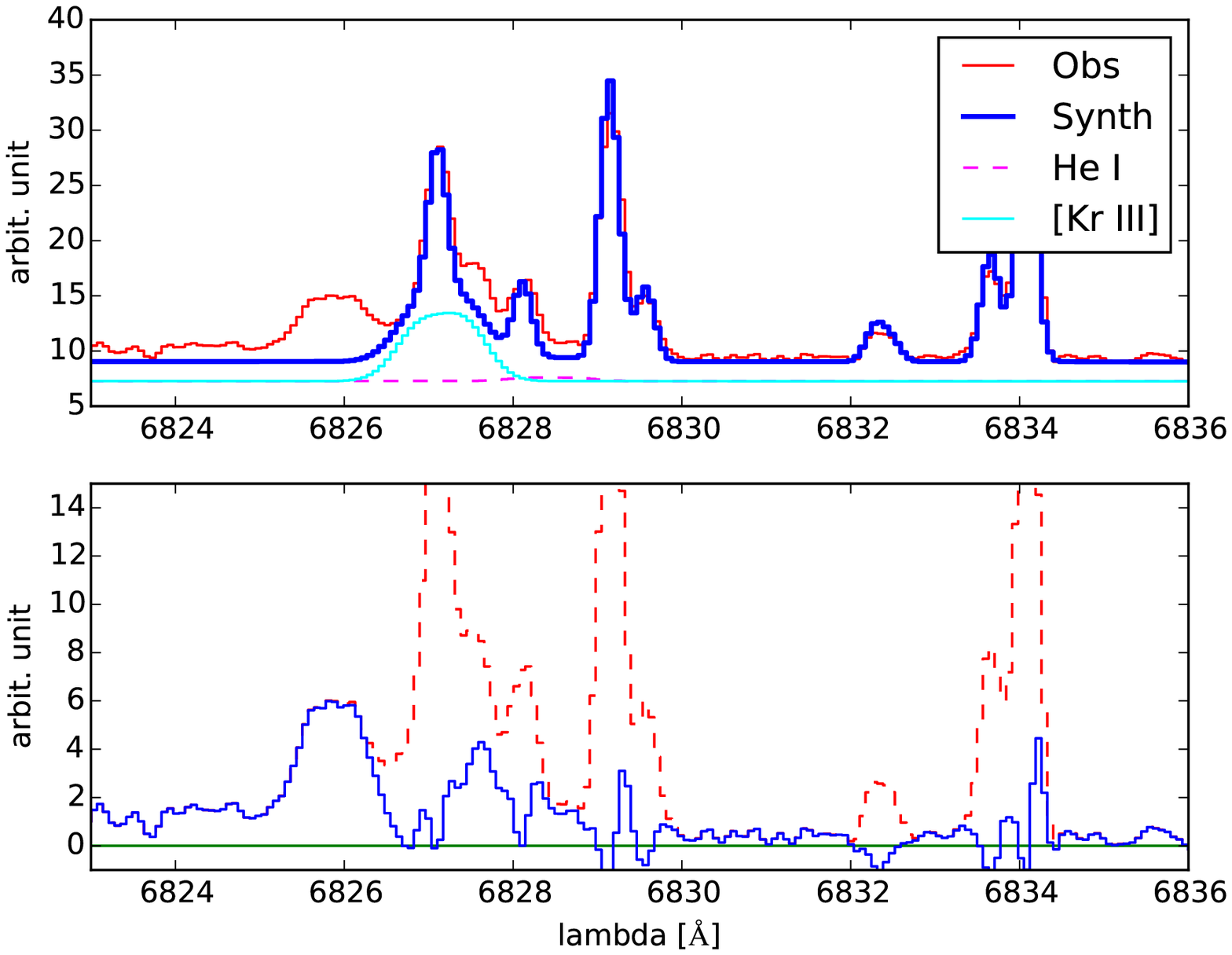}
\caption{PySSN fitting of the {\fkriii} $\lambda$6826.70 + {\hei} $\lambda$6827.88 lines in the spectra of He\,2-86 (top panel), PC\,14 (middle panel) and Pe\,1-1 (lower panel). Colors and labels are the same than in Figure~\ref{kriii_XSSN}. The feature labeled as CT is a feature produced by charge transfer coming from the emission of {\ha} in a contiguous echelle order.}
\label{kriii_XSSN_GR12}
\end{center}
\end{figure}

Nucleosynthesis models also predict that neutron-capture elements correlate with the C/O ratio, as carbon is brought to the surface of AGB stars along with {\emph s}-processed material during third dredge-up episodes \citep{ bussoetal01, gallinoetal98, karakaslattanzio14}.
In the upper panel of Figure~\ref{KrO_CO} we show the correlation found by \citet{sterlingdinerstein08} for PNe of their sample with known abundances of C (from UV lines) and their Kr abundances obtained from near IR {\fkriii} line detections (red squares). This correlation is marginal due to the large uncertainties in the Kr/O and C/O ratios. 
We have overplotted our result for NGC\,3918 from optical data for both the C/O and the Kr/O ratios (big blue dot). We adopt C/O=0.87$\pm$0.25 from {\cii} and {\oii} recombination lines \citep[see~Fig~\ref{oii_lines} and the ICFs given by ][]{kingsburghbarlow94}. C/O ratios from ORLs generally agree with those derived from UV/optical CELs \citep{yliuetal04b, delgadoingladarodriguez14} and, additionally, they are not affected by aperture corrections and extinction uncertainties, which can potentially affect C/O ratios derived from CELs. 
Additionally, we also include some data from \citet{garciarojasetal09} and \citet{garciarojasetal12} where Kr lines are detected. From \citet{garciarojasetal12} only NGC\,6369 has detections of more than one ionization stage of Kr ({\fkriii} and {\fkriv} lines are detected). For the other objects, only {\fkriv} lines are reported. 
However, a reinspection of the spectra in \citet{garciarojasetal12, garciarojasetal13} has brought to light detections of the {\fkriii} $\lambda$6826.70 line in the spectra of He\,2-86, PC\,14, Pe\,1-1 and, possibly NGC\,2867. These lines are not reported in \citet{garciarojasetal09} and \citet{garciarojasetal12} because they are strongly affected by charge transfer and telluric emission features. We use PySSN to deblend this line from the charge transfer feature, the telluric emission lines and a close {\hei} line (see Fig.~\ref{kriii_XSSN_GR12}). The fluxes we obtain are shown in Table~\ref{new_fluxesGR12}. We consider in all cases that the uncertainties in the flux measurements are larger than 40\%.
Then, to compute the total Kr abundance for these new objects we use  Eq. (3) of \citet{sterlingetal15} when only {\fkriv} lines are detected (M\,1-61 and NGC\,2867) and Eq. (4) when both {\fkriii} and {\fkriv} lines are detected (He\,2-86, NGC\,6369, PC\,14, Pe\,1-1). We computed C/O ratios by using the C$^{2+}$ and O$^{2+}$ abundances reported by \citet{garciarojasetal09} and \citet{garciarojasetal13} for these objects and applying the ICF recipes given by \citet{kingsburghbarlow94} (green triangles) in order to be consistent with the C/O ratios derived by \citet{sterlingetal07}. Additionally, we add two objects (M\,1-25 and M\,1-32) for which \citet{sterlingetal07} report [Kr/(O,Ar)] ratios, but reanalyze the data including our log(C/O) ratios derived from C$^{2+}$/H$^+$ and O$^{2+}$/H$^+$ from ORLs reported by \citet{garciarojasetal13} and the ICFs by \citet{kingsburghbarlow94} (yellow squares).

\setcounter{table}{15}
\begin{table}
\centering
\caption{I({\fkriii})/I({\hb}) $\lambda$6826.70 line ratios (I(H$\beta$) = 100) from the \citet{garciarojasetal12} PNe sample.}
\label{new_fluxesGR12}
\begin{tabular}{lc}
\hline
Object  & $I(\lambda)/I({\rm H}\beta)$ \\
\hline
He\,2-86 & 0.0104 \\
NGC\,6369 & 0.0210  \\
PC\,14 & 0.0229  \\
Pe\,1-1 & 0.0436  \\
\hline
\end{tabular}
\end{table}

\citet{sterlingetal15} find that Eq. (3) seems to produce larger Kr abundances than other equations, especially for low-ionization objetcs, and they do not recommend to use this equation when {\fkriii} emission is clearly detected. In Table~\ref{ICFs_Kr} we show the total Kr abundances obtained from all equations of \citet{sterlingetal15} for NGC\,3918 and from Eqs. (2), (3) and (4) for the objects of the \citet{garciarojasetal09} and \citet{garciarojasetal12} sample. Our results are very similar to those of \citet{sterlingetal15} and Eq. (3) gave a very high abundance compared to other equations. The exception is NGC\,3918, owing to its high ionization.

\setcounter{table}{16}
\begin{table*}
\caption{Comparison of Kr abundances computed with the ICFs by \citet{sterlingetal15}} 
\label{ICFs_Kr} 
\begin{tabular}{lcccccc} 
\hline\hline 
PN 	& \multicolumn{6}{c}{12+log(Kr/H)}$^{\rm a}$  	\\
          & ICF Eq. (1)       & ICF Eq. (2)        & ICF Eq. (3)        & ICF Eq. (4)        & ICF Eq. (5)        & ICF Eq. (6)         \\
\hline 
NGC\,3918  & 3.80$^{+0.07}_{-0.11}$ & 3.86$\pm$0.06 & 3.83$\pm$0.05 & 3.92$^{+0.07}_{-0.04}$ & 3.85$\pm$0.04 & 3.92$\pm$0.05	\\
He\,2-86    &  3.64$\pm$0.15 &  3.67$\pm$0.17 &   4.08$\pm$0.07  &3.62$\pm$0.07   &        ---        &    ---       \\
M\,1-61     & 	---		 & 	---		&   3.78$\pm$0.10  &	     ---           &        ---        &    ---       \\
NGC\,2867  & 	---		 & 	---		&   4.35$\pm$0.04  &	     ---           &        ---        &    ---       \\
NGC\,6369  & 3.92$\pm$0.11& 3.48$\pm$0.14& 4.52$\pm$0.06 & 3.83$\pm$0.07 &        ---        &    ---       \\
PC\,14         & 3.93$\pm$0.15& 3.71$\pm$0.16& 4.37$\pm$0.09 & 3.81$\pm$0.08 &        ---        &    ---       \\
Pe\,1-1       & 3.95$\pm$0.15 & 3.97$\pm$0.16 & 4.64$\pm$0.05 & 3.96$\pm$0.09  &        ---        &    ---       \\
\hline
\end{tabular}
\begin{description}
\item[$^{\rm a}$] ICFs Eqs. (1) and (2) based on Kr$^{2+}$ abundances; ICF Eq. (3) based on Kr$^{3+}$ abundances; ICF Eq. (4) based on Kr$^{2+}$ and Kr$^{3+}$ abundances;  ICF Eq. (5) based on Kr$^{3+}$ and Kr$^{4+}$ abundances;  ICF Eq. (6) based on Kr$^{2+}$, Kr$^{3+}$, and Kr$^{4+}$ abundances.
\end{description}
\end{table*}

In the lower panel of Figure~\ref{KrO_CO} we only include the objects with {\fkriii} and {\fkriv} lines detected and compute the C/O ratios using {\cii} and {\oii} recombination lines and the ICFs given by \citet{delgadoingladaetal14}. We  overplot our result for NGC\,3918  adopting C/O=1.00$\pm$0.29 from {\cii} and {\oii} recombination lines and the ICFs given by \citet{delgadoingladaetal14}, which is in excellent agreement with the result obtained from the sum of ionic abundances (C/O=1.02$\pm$0.12). 
A least squares linear fit taking into account uncertainties in both axis yields:

\begin{equation}
[{\rm Kr}/{\rm O}] = (0.480 \pm 0.027)  + (0.901 \pm 0.240) \log ({\rm C}/{\rm O}),
\end{equation}

\noindent where the uncertainties include both the dispersion and the individual uncertainties for each object. The Spearman coefficient for the two variables is $r$(Spearman) =0.94$\pm$0.06 showing clearly that both quantities follow a monotonic relation. The Pearson coefficient is somewhat lower, but is also close to 1, $r$(Pearson)=0.86$\pm$0.07, indicating that the relation is probably linear; however, these values should be taken with caution given the low number of points in the correlation. Adding more points to the sample will strengthen even further our conclusions. It is important to remark that this fit is very similar to the ones obtained by \citet{sterlingdinerstein08} and \citet{sterlingetal15} but with a much better correlation coefficient.
 
As can be shown, our approach of utilizing multiple Kr ions to derive Kr abundances and determining C/O ratios from ORLs gives gives more precise results than the combined UV and near-IR data, and minimizes the uncertainties given by aperture effects in the different wavelength ranges as well as the uncertainties related to extinction. Additional very deep optical observations of a large sample of PNe could help to confirm this correlation of [Kr/O] with C/O (and also of [Br/O], [Se/O], [Rb/O] and [Xe/O]) to constrain nucleosynthesis models.

\begin{figure}[h]
\center
\includegraphics[width=9cm]{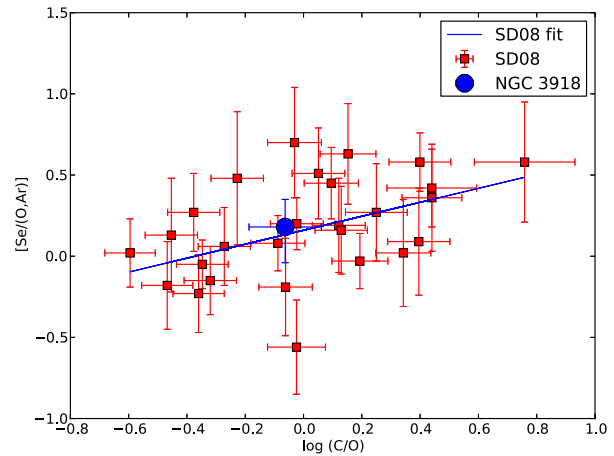}
\caption{[Se/(O,Ar)] $vs.$ log(C/O). Red squares represent the data presented by \citet{sterlingdinerstein08} where C/O ratios were computed from UV  forbidden lines. The blue dot represent the result obtained for NGC\,3918. For NGC\,3918 the C/O ratio was computed from optical recombination lines and using the ICF scheme by \citet{kingsburghbarlow94}. The least squares fit to the \citet{sterlingdinerstein08} data is shown as a blue line.}
\label{SeO_CO}
\end{figure}

Similarly to what we have done for Kr, in Figure~\ref{SeO_CO} we show the correlation found by \citet{sterlingdinerstein08} for a sample of PNe with known abundances of C (from UV lines) and Se abundances obtained from near IR {\fseiii} line detections (red squares). The least squares fit to these data is shown as a blue line. We overplot the point for NGC\,3918 where the C/O ratio was computed from optical recombination lines and using the ICF scheme by \citet{kingsburghbarlow94}. Our result is consistent with the fit found by \citet{sterlingdinerstein08}. 

\section{Summary}\label{conclu}

In this work we develop a detailed study of the physical conditions and the chemical content of the planetary nebula NGC\,3918, based on deep, high-resolution (R$\sim$40000) spectrophotometric data obtained with the UVES echelle spectrograph at the VLT. We identify and measure more than 750 emission lines, making ours one of the deepest spectra ever taken for a planetary nebula. 

Thanks to the high ionization of NGC\,3918, we detect CELs of multiple ionization stages. In addition to the classical CELs of N, O, Ne, S, Cl, and Ar, we detect a large number of CELs of refractory element (Ca, K, Cr, Mn, Fe, Co, Ni and Cu).  We also detect very faint CELs of several neutron-capture elements (Se, Kr, Rb, and Xe). ORLs of multiple ionization stages of C, N, O, and Ne are also detected in our spectrum.

We compute physical conditions from a large number of diagnostics, which are highly consistent among themselves assuming a three-zone ionization scheme.

We compute elemental abundances by using a state-of-the-art ICF scheme or simply adding ionic abundances. When multiple ionization stages are detected, both abundance determinations are in very good agreement, demonstrating the quality of the recent ICF scheme for high ionization planetary nebulae. 
We also determine C/O ratios from pure recombination lines. These ratios derived from {\cii} and {\oii} ORLs and ICFs are consistent with those derived from the direct sum of all the ionic species detected, showing that C/O ratios from ORLs have a clear advantage over UV determinations

Several refractory elements lines (Ca, K, Cr, Mn, Co, Ni and Cu)  were detected in the spectrum of NGC\,3918. We estimate the abundance of Ca and K making use of ICFs derived from simple photoinization models. For several Fe-peak species we could not compute abundances owing to the lack of atomic data. New atomic data (specifically, transition probabilities
and effective collision strengths for transitions between the lowest ~50 energy levels) for these species are urgently needed to compute gas-phase abundances of these species and provide valuable information about the dust grain chemistry in PNe and AGB winds.

Chemical abundances of Se, Kr, Rb, and Xe are computed, in the case of Kr with unprecedented accuracy, thus constraining the efficiency of the {\emph s}-process and convective dredge-up in NGC\,3918's progenitor star. We find that Kr is strongly enriched in NGC\,3918 and that Se is less enriched than Kr, in agreement with the results of previous papers and with predicted {\emph s}-process nucleosynthesis. As we compute only lower limits to Rb and Xe enrichments we cannot draw definitive conclutions, but the results suggest that Xe is not as enriched by the {\emph s}-process in NGC\,3918 as is Kr and, therefore, that neutron exposure is typical of modestly subsolar metallicity AGB stars. The modest Rb enrichment relative to Kr and Xe seem to indicate that the progenitor mass of NGC\,3918 is between 1.5 M$_{\odot}$ and 5--6 M$_{\odot}$. 

A clear correlation is found when representing [Kr/O] $vs.$ log(C/O) for NGC\,3918 and other objects with detection of multiple ions of Kr in optical data confirming that carbon is brought to the surface of AGB stars along with {\emph s}-processed material during third dredge-up episodes, as predicted by nucleosynthesis models. Additional deep high-resolution spectrophotometric data are needed for [Br/O], [Se/O], [Rb/O] and [Xe/O] $vs.$ C/O to put narrow constraints to nucleosynthesis models.

\section*{Acknowledgments}

This work is based on observations collected at the European Southern Observatory, Chile, 
proposal number ESO 090.D-0265(A). This work has been funded by the Spanish Ministry of Economy and Competitiveness (MINECO) under the grant AYA2011-22614, CONACyT, Mexico under grant CB-2010/153985 and by the DGAPA-UNAM, Mexico under grant PAPIIT-107215. We want to thank the annonymous referee for his/her valuable comments. JGR acknowledges support from Severo Ochoa excellence program (SEV-2011-0187) postdoctoral fellowship. VL acknowledges support of the Spanish Ministry of Science and Innovation through grant AYA 2011-22614. NCS gratefully acknowledges support of this work from an NSF Astronomy and Astrophysics Postdoctoral Fellowship under award AST-0901432 and from NASA grant 06-APRA206-0049. GDI gratefully acknowledges support from a DGAPA postdoctoral grant from the Universidad Nacional Autónoma de México (UNAM) and the GSI Foundation. JGR acknowledges all staff, employees and guests of the Instituto de Astronom\'{\i}a at UNAM, where part of this work was done. We also thank C. Esteban, D.~A. Garc\'{\i}a-Hern\'andez, D. P\'equignot, and M. Rodr\'{\i}guez for fruitful discussions.


\appendix

\onecolumn
\setcounter{table}{1}
\begin{longtable}{ccccccccc} 
\caption{Observed and reddening corrected line ratios (F(H$\beta$) = 100) and line identifications in NGC\,3918.\label{lineid}}\tabularnewline
\hline\hline 
$\lambda_0$ (\AA\ ) & Ion & Mult. & $\lambda_{\rm obs}$ & $V_{\rm rad}$ (km s$^{-1}$) & $F(\lambda)/F({\rm H}\beta)$$^{\rm a}$ & $I(\lambda)/I({\rm H}\beta)$$^{\rm b}$ & Err(\%) & notes \tabularnewline
\hline
\endfirsthead
\caption{continued.}\tabularnewline
\hline\hline 
$\lambda_0$ (\AA\ ) & Ion & Mult. & $\lambda_{\rm obs}$ & $V_{\rm rad}$ (km s$^{-1}$) & $F(\lambda)/F({\rm H}\beta)$$^{\rm a}$ & $I(\lambda)/I({\rm H}\beta)$$^{\rm b}$ & Err(\%) & notes \tabularnewline
\hline
\endhead
\hline
\endfoot
\endlastfoot
 3109.18 &        {\fariii} &        3P-1S & 3108.89 & -27.97 &   0.485 &   0.648 &	 9 &	 \tabularnewline
 3115.67 &          {\oiii} &    3S-3P$_0$ & 3115.42 & -24.06 &   0.165 &   0.219 &	11 &	 \tabularnewline
 3118.61 &         {\fcliv} &        3P-1S & 3118.42 & -18.28 &   0.104 &   0.138 &	14 &    \tabularnewline
 3121.63 &          {\oiii} &    3S-3P$_0$ & 3121.41 & -21.13 &   2.444 &   3.249 &	 8 &	 \tabularnewline
 3132.79 &          {\oiii} &    3S-3P$_0$ & 3132.58 & -20.09 &  53.51 &  70.93 &	 8 &	 \tabularnewline
 3187.84 &           {\hei} &            3 & 3187.50 & -31.99 &   2.492 &   3.255 &	 8 &	 \tabularnewline
 3203.10 &          {\heii} &          3.5 & 3202.89 & -19.68 &  13.62 &  17.72 &	 8 &	 \tabularnewline
 3218.19 &          {\neii} &    4D$_0$-4F & 3217.95 & -22.36 &   0.0845 &   0.1096 &	16 &	 \tabularnewline
\end{longtable}

\clearpage

\twocolumn


\end{document}